\documentclass[onecolumn]{emulateapj}
\usepackage{epsfig}
\usepackage{amsmath}
\usepackage{graphicx}
\usepackage{rotating}
\usepackage{tabularx}
\usepackage{changepage}
\usepackage{amssymb}
\usepackage{subfigure}

\shorttitle{Imaging Redshifts for BL Lacs}
\begin{document}

\title{Imaging Redshift Estimates for BL Lacertae Objects}
\author{Aaron M. Meisner \& Roger W. Romani}
\affil{KIPAC/Dept. of Physics, Stanford University, Stanford, CA 94305-4060}
\email{ameisner@stanford.edu, rwr@astro.stanford.edu}
%\date{\today}
%\maketitle

\begin{abstract}
	We have obtained high dynamic range, good natural seeing $i^\prime$ images
of BL Lacertae objects (BL Lacs) to search for the AGN host and thus constrain 
the source redshift. 
These objects are drawn from a sample of bright flat-spectrum radio sources
that are either known (via recent {\it Fermi} LAT observations) gamma-ray emitters
or similar sources that might be detected in continuing gamma-ray observations. 
All had spectroscopic confirmation as BL Lac sources, but no redshift 
solution. We detected hosts for 25/49 objects. As these galaxies have been 
argued to be standard candles, our measured host magnitudes provide redshift
estimates (ranging from 0.2--1.0).  Lower bounds are established on the redshifts
of non-detections. The mean of the fit redshifts (and lower limits) is higher
than those of spectroscopic solutions in the radio- and gamma-ray- loud
parent samples, suggesting corrections may be needed for the luminosity function
and evolution of these sources.
\end{abstract}


\section{Introduction}
	
	BL Lac objects, being highly continuum-dominated, are a perennial
problem for those wishing to study the evolution of AGN populations. The lack
of visible broad lines and, even more, the low equivalent width of host
absorption features makes redshift determinations extraordinarily difficult.
Yet, since the continuum domination is a manifestation of the good alignment
of the relativistic jet outflow to the Earth line-of-sight, \citep{up95} % Urrey and Padovani
knowledge of the distance, and hence luminosity scale, of these sources 
is very interesting.

	This problem has become particularly important with gamma-ray detection
of large numbers of blazars with the {\it Fermi} LAT \citep{BSC}. It has been known
since {\it EGRET} that radio loud flat spectrum blazars dominate the
bright extragalactic sources in the GeV sky \citep{ha99,ma01,seet05}; these objects
are flat-spectrum radio quasars (FSRQ) and radio loud BL Lacs. We have good
techniques in place to identify likely radio counterparts for the gamma-ray sources.
Extensive spectroscopy campaigns, especially on the EGRET-like `CGRaBS'
sample \citep{het08} have provided nearly complete
redshifts and characterizations of the FSRQ. However, despite extensive observation,
including 8+m telescope integrations, less than half of the BL Lac
counterparts have spectroscopic IDs. In some cases, high quality spectroscopy
can give useful lower limits on the source redshift \citep{shaw09}, but many 
remain unconstrained.
In the first LAT blazar catalog \citep{ab09a}, the problem is even more pronounced
since the excellent high energy response of the LAT favors detection of hard
spectrum sources. BL Lacs are substantially harder ($\langle \Gamma \rangle \approx 2.0$)
than FSRQ ($\langle \Gamma \rangle \approx 2.4$) and so provide a larger fraction,
$\sim40$\% of the {\it Fermi} blazar sample, than was seen by {\it EGRET}. 

	We attempt here to constrain the redshifts of radio loud BL Lacs which have 
shown no convincing spectroscopic redshift solution, despite sensitive 
observations on large telescopes, by high dynamic range imaging searches for the AGN host. 
Our targets are drawn from the `CRATES' catalog of bright flat-spectrum $|b|>10^\circ$ 
radio sources \citep{het07}. Specifically, in this program we targeted the subset of
CRATES consisting of $i$) sources selected as likely gamma-ray blazars before 
the Fermi mission (i.e. CGRaBS sources, Healey et al. 2008) and $ii$) sources
that are likely counterparts to LAT sources detected early in the mission
\citep{ab09a,1FGL}. All sources are at Declination $> -20^\circ$ and all have
been shown to display featureless optical spectra \citep{het08,shaw09,shaw10},
with sensitive (4m+ class) observations.  We were able to image $\sim 2/3$ of the 
sources satisfying these critera at the time of the observing campaigns.

	It has been claimed \citep{sba05} that BL Lac host galaxies detected with 
HST imaging are remarkably 
uniform giant ellipticals with $M_R=-22.9$; accordingly host detections can give 
redshift estimates and upper limits on the host flux can give lower limits on the 
distance. Here we do not test this assumption or the possible biases that would
select a modest magnitude range for the detected host sample. We simply apply the 
method to extract redshift constraints, noting that while individual
redshifts are doubtless imprecise, the estimates can still be useful for statistical
studies of BL Lac evolution and as a guide to and comparison with other methods of redshift
estimation. Conversely, when precision spectroscopic redshifts become available,
our measurements can be used to help constrain the host evolution. 

	Since we will be searching for de Vaucouleurs profile excesses in the
wings of the stellar PSF of the bright BL Lac core, we require good natural seeing
and a moderate field of view for adequate comparison stars. For example, while
near-IR adaptive optics can deliver superior PSF cores, the wings of the source
at $>$90\% encircled energy are extensive and often quite variable over the
small corrected FoV. This prevents the accurate PSF modeling and subtraction 
required to obtain host measurements whose integrated magnitude can be 10\% or
less of the core flux.

\section{Observations \& Data Reduction}

All observations were carried out using the WIYN 3.6m telescope at Kitt 
Peak National Observatory, which features good natural seeing and a $\sim 10^\prime$
FoV with the standard mosaic imagers.

\subsection{OPTIC Camera}

	We employed OPTIC on the nights of 2008 Oct. 31 -- Nov. 2 for our initial
imaging run. This 2$\times$2k$\times$4k camera covers 9.6$^\prime$ at
0.14$^{\prime\prime}$/pixel and features Lincoln Labs orthogonal transfer
CCDs, which allow rapid electronic (OT) tracking using an appropriate in-field guide star.
With a relatively short (25s) read-out time, we were able to combine short
exposures that avoided core saturation with longer exposures that probe for
the host galaxies in the PSF wings. The targets were imaged with 
the SDSS $i^\prime$ filter KPNO 1586. We observed 31 targets, with total
on-source time ranging from 210\,s to 2100\,s (Table \ref{optic_results}).
Sky conditions were photometric through the run, and we were able to
employ OT guiding for the bulk of the sources, with correction rates
up to 50 Hz. The final $i^\prime$
image quality had a median FWHM of 0.65$^{\prime\prime}$; the best 25\%
of the sources had image stacks with FWHM $< 0.5^{\prime\prime}$. 
A period during which mirror cooling was lost resulted in some observations
with poor PSF (up to 1$^{\prime\prime}$). Software problems in addition
cost several hours.

	While OT guiding significantly improves the source PSF, which
is crucial for our science goals, it does create some complications during
data reduction. In particular, flat-field frames must be created for each
individual exposure, with weighted exposure per pixel following the
history of OT offsets during each individual exposure. This creates some
challenges in assembling the final mosaic images, as defects can
affect adjacent pixels. The `smoothing' of the flat-field response
also affects the noise statistics of the final image. The data were
processed using routines in the IRAF package \verb|mscred|. 
With the excellent PSF (as small as 0.4$^{\prime\prime}$) achieved 
for some frames, we required superior camera distortion
corrections at each position angle; for some orientations a lack
of field stars limited the accuracy of the astrometric solution.	
After registration and exposure-weighting, median combined image
stacks were prepared for further analysis.

\subsection{Mini-Mosaic Camera}

	Observations were made using the Mini-Mosaic camera (MiniMo) on the nights of 2009 
March 24-25, under highly variable conditions. The 2$\times$2k$\times$4k mosaic 
similarly has a plate scale of 0.14$^{\prime\prime}$/pixel and covers a field  
of 9.6 arcmin.
We again employed the SDSS $i^\prime$ filter. Because of the 
long ($\sim$3 minute) readout time of the Mini-Mosaic camera, most objects had
3-5 dithered exposures of 300s each. These relatively long exposures meant that
many of the BL Lac targets had saturated cores, despite the typically poorer final
image quality during this run (median final FWHM 0.84$^{\prime\prime}$).
The exposure sequences and final DIQ for each object are listed in Table 
\ref{minimo_results}. In all, 18 BL Lacs were observed. Since early LAT detections
had been announced by this run \citep{BSC} we were able to specially target
known gamma-ray emitters lacking redshift solutions. The sky transparency 
was variable and half the run was lost to high winds and poor seeing.

	MiniMo employs conventional CCDs, and so standard reductions were 
conducted with the IRAF package \verb|mscred|. Camera distortions were mapped
using USNO A2.0 catalog stars; these were also used to assign a WCS to each frame.
After processing and bad pixel correction the individual dither frames
were aligned using unsaturated stars near the BL Lac,
exposure weighted and median combined using standard sigma-clipping algorithms
to produce our final images. MiniMo suffers from `ghost' images due to
amplifier cross-talk. Care was taken during target acquisitions to ensure
that these did not fall in the vicinity of the BL Lac. We made no attempt to
correct these ghosts, but avoid using any affected comparison stars.

\section{Calibration}

	A number of the target fields were covered by the SDSS, so we were able to
establish our zero point directly from standard aperture photometry 
of unsaturated field stars (16 $\lesssim m_{i^\prime} \lesssim$ 19).
For the OPTIC run only six fields had suitable stars. The median rms of these
zero point measurements was 0.052 mag. While this was significantly larger
than the intrinsic SDSS photometric errors, the scatter is smaller than
our typical host photometric errors and much smaller
than the scatter in the claimed absolute magnitude of the BL Lac host ($\delta M=0.5$\,mag). For 
the exposures lacking SDSS comparison stars, we applied the mean zero point
after correction for atmospheric extinction \citep{massey02}. Given that
conditions were photometric throughout, we are conservative in assigning
the 0.068 scatter between the zero point measurements as the final zero-point error.
%Six fields (J0909+0200, J0050$-$0929, J0502+1338, J1927+6117, J2050+0407, J2200+2137) 

	While conditions were variable during the MiniMo observations, 12/18
fields had stars suitable for direct SDSS cross-calibrations. In particular
for four observations logged as cloudy we were able to establish
in-field calibration. Taking the rms zero point from the remainder of the fields,
we established, after extinction correction, a zero point with scatter
of 0.036 mag. This was applied to the remaining target fields.

\section{PSF Modeling}

	Accurate PSF modeling is essential to extract the BL Lac host from the wings
of the nuclear point source. We built an effective PSF in the stacked image
of each BL Lac using the IRAF/DAOPHOT \verb|psf| routine. PSF stars were selected 
from those that appeared in all sub-frames making up the image stack, had stellar
FWHM and appeared isolated. PSF stars were selected by hand and were individually checked
for consistency with a stellar FWHM. In practice the process was iterative; some otherwise
desirable PSF template stars had faint unresolved sources in the wings, so we modeled
and subtracted these sources before recombining for an improved PSF. Point sources
as faint as $i^\prime \sim 25$ were removed in this manner.  Each template 
PSF has a radius of 5.6$^{\prime\prime}$ and covers over 11 FWHM even for our worst DIQ images. 
The background level for each PSF star was determined from the mode of a 2$-$4$^{\prime\prime}$
wide annulus starting at 7$^{\prime\prime}$ from the star. In a few of the images
with poor seeing the background was determined from larger radius, to ensure insignificant
contamination from the PSF wings.
To best model the PSF wings, we also used some stars with nonlinear cores. Above the
non-linearity level (30,000 ADU for OPTIC, 40,000 ADU for MiniMo), the cores were masked
in the PSF stack. 

\begin{figure}
\centering
\mbox{\subfigure{\includegraphics[width=3.5in]{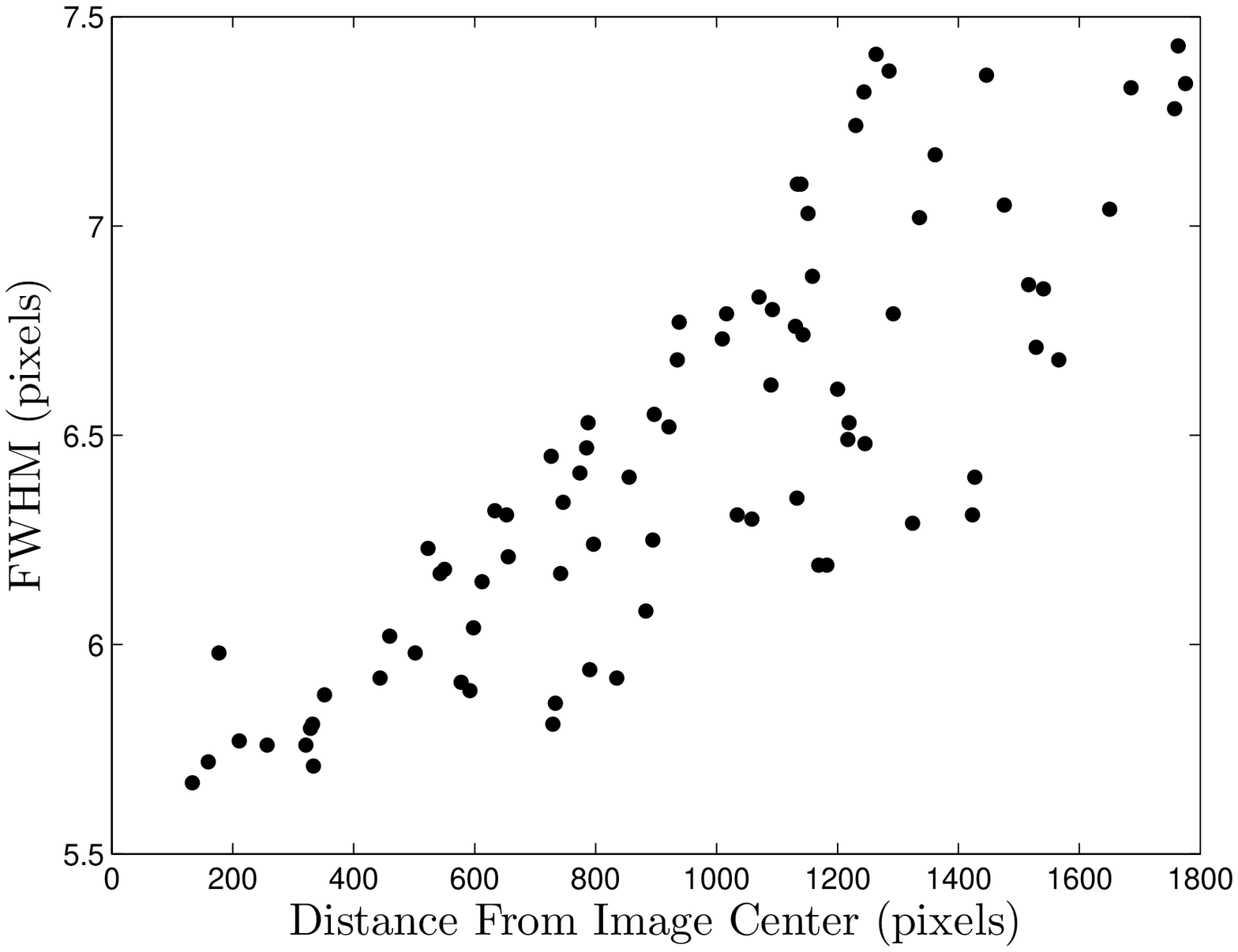}}\quad
\subfigure{\includegraphics[width=3.5in]{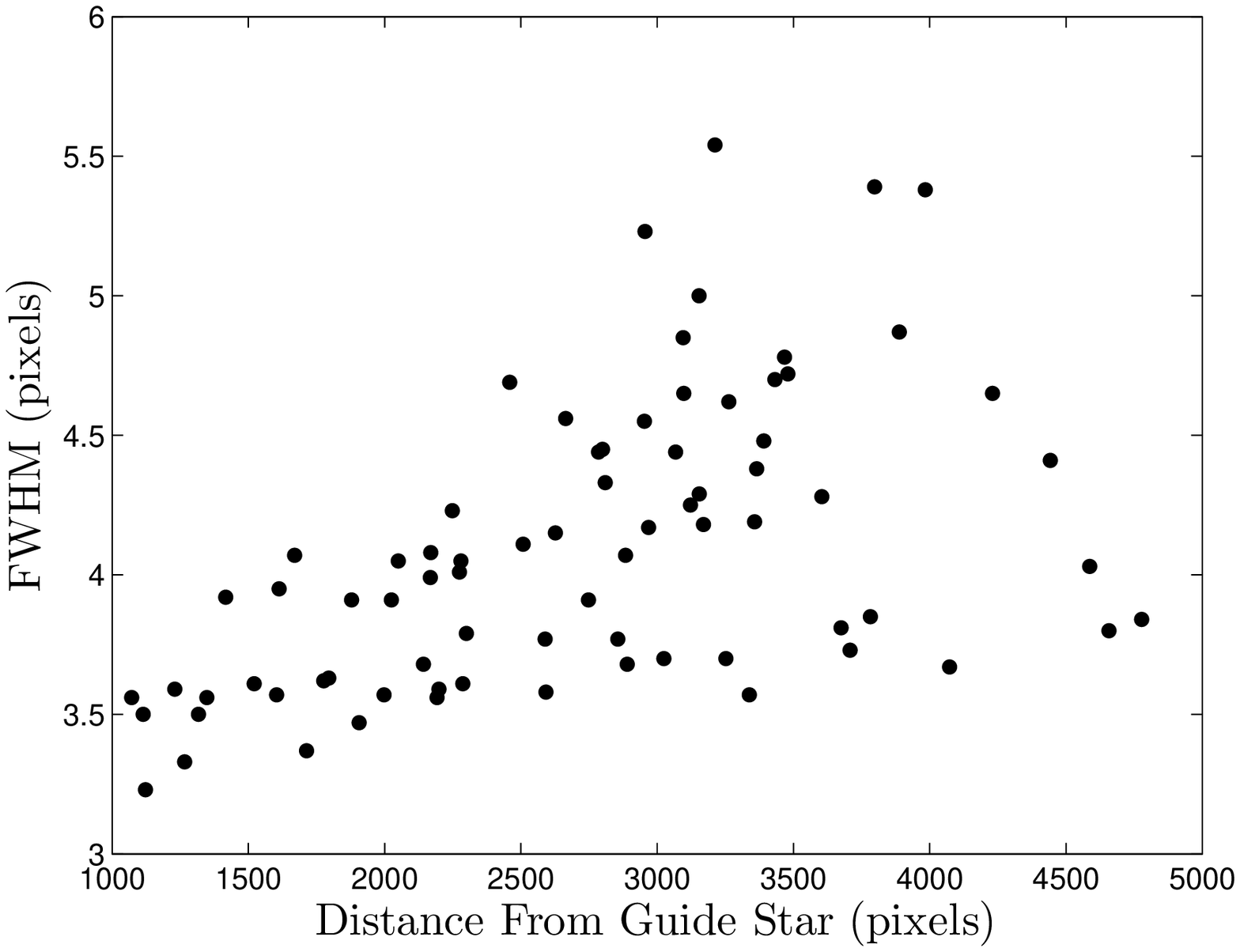} }}
\caption{ PSF variation across the field for OPTIC data. Left: FWHM variation 
with distance from the field center in the image of J0110+6805 (observed without OT guiding).
Right: variation with offset from the star used for rapid guiding near J0612+4122. 
} \label{fig:comparison}
\end{figure}

	We found that a PSF model with linear gradients across the image \citep{davis94} was
required to take full advantage of the small FWHM of the best OPTIC data. To guide our
fitting we examined the spatial dependence of the FWHM. Three terms might be expected
to contribute to PSF degradation. First, we expect the efficacy of the low order 
atmospheric corrections to fall off with distance from the guide star. We attempted to
mitigate this effect by guiding in the target quadrant, whenever possible. Second,
distortions in the camera system are expected to increase away from the optical axis
(assumed to be near the field center). Finally, since our principal frame registration
is on the BL Lac core, registration or residual rotation errors could degrade the PSF
as a function of distance from the target itself. To test for these effects, we collected
FWHM measurements of unresolved sources across a number of fields. In no case did we find
the PSF was primarily correlated with distance to the target either in the individual
exposures or in the final combined images, implying that stacking error was not
important. In some cases the PSF degradation appeared correlated with
distance from the optical axis, implicating camera distortions, but for other
images the distance from the guide star appeared to dominate (see Figure \ref{fig:comparison}).

	With the poorer seeing experienced during our MiniMo run, we found that the spatial
variation of the PSF was much less important, with the FWHM varying by no more than
a few $\sim 0.1$\,pixels ($\sim0.03^{\prime\prime}$) across the field of view. Again stacking errors
were negligible. However for consistency we fit for and applied linear gradients in the
PSF model.

	Thus, in forming the final model PSF for the analysis, we were able to draw
stars from across the fields avoiding only the outermost corners of the frames, where there
was notable degradation. In general between $\sim$ 1/6 and $\sim$ 1/5 of the PSF stars used 
were saturated in their cores (only $\sim$ 1/9 with MiniMo, given the poorer seeing). 
Tables \ref{optic_results} and \ref{minimo_results} list the number of PSF stars used in each 
image. The median number of PSF stars for each field was 30 for both OPTIC and MiniMo.

\section{Host Galaxy Model Fitting }

	Previous studies \citep{urry00p1, urry00p2, urry05} indicate that BL Lac hosts 
are better fit by de Vaucouleurs profiles than exponential disks. 
We therefore adopt simple de Vaucouleurs profiles with fixed Sersic index 4,
convolving the model with the image PSF, in our galaxy fitting. We did find examples 
of disturbed galaxies and hosts with faint companions, but no host appeared disk-dominated.
Our BL Lac image model includes up to five parameters, introduced hierarchically. We start by 
fitting for nuclear (PSF) flux plus possible host. We assume that the host galaxy isophotes are centered on 
the nucleus \citep{urry00p3}, and fit or bound the host flux, initially assuming a a fixed
effective radius of $R_{\rm e}=1.64^{\prime\prime}$ (10\, kpc at $z=0.5$; we assume 
$\Omega_{m} = 0.3$, $\Omega_{\Lambda} = 0.7$ and $H_{0} = 70 $ km s$^{-1}$ Mpc$^{-1}$
throughout this paper). We next allow the de Vaucouleurs profile $R_{\rm e}$ to vary, 
when required by a high significance for a measurement of $R_{\rm e}$ in the fit.
Finally for a few
BL Lacs, we find a significant detection of host ellipticity; in these cases we
include the ellipticity $\epsilon$ and position angle $\theta$ as additional free parameters in the 
fit. After fitting the host normalization, we report galaxy magnitudes by extrapolating 
the integrated light to infinity.

	To prepare for host fitting, we determine the centroid position of the core PSF,
using the DAOPHOT task \verb|peak|, which matches to the model PSF, excluding pixels
above our (conservative) non-linearity threshold (30,000 ADU for OPTIC, 40,000 ADU for MiniMo), and
delivering centroid coordinates accurate to $\pm 0.01$ pixels. Since a number of our
images were saturated, this clipping was important for obtaining accurate centroids.
In almost all cases, the galaxy surface brightness dropped below the background
surface brightness fluctuations within our standard $5.6^{\prime\prime}$ PSF radius.
Accordingly the BL Lac image model was fit to $11.2^{\prime\prime} \times 11.2^{\prime\prime}$ 
cutouts from the image stack centered on the BL Lac core pixel. DAOPHOT provides a 
2$\times$-oversampled
PSF model. This was re-binned to the original pixel size, using the precisely determined centroid
and then used in the subsequent fitting. This PSF centering and re-sampling
proved to be quite accurate -- dipole structure was never evident in residuals
to our final BL Lac model, whereas with even 0.05 pixel shifts for the BL Lac core
position such residuals were obvious.

\begin{figure} [h]
\begin{center}
\includegraphics[scale=0.65]{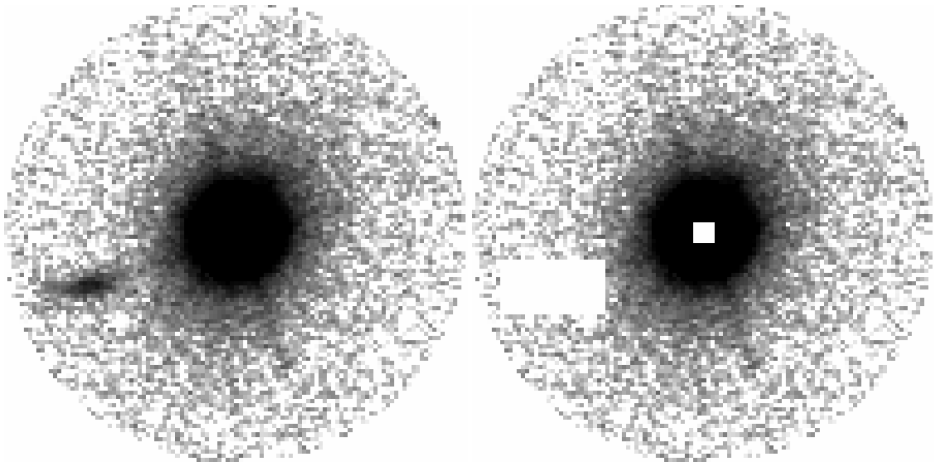}
\end{center}
\caption{\label{fig:J0814_blanking}Preparing the fitting region for J0814+6431 (image
FWHM $0.45^{\prime\prime}$). One 
resolved companion and pixels near the centroid of the nuclear component
are masked. The corrected FOV is $7.7^{\prime\prime}$ in radius; N up, E left.}
\end{figure}

	We apply our BL Lac fit to a subset of the pixels in the cutouts around each
BL Lac. First all pixels beyond $5.6^{\prime\prime}$ (outside the PSF model) are masked.
Pixels associated with resolved objects within the fitting radius but more
than three FWHM from the BL Lac core are also masked (in practice we exclude rectangular regions,
see Figure \ref{fig:J0814_blanking}). 
%% check this 3 FWHM number with AM
Unresolved objects more than three FWHM away are fit to the PSF model and
then treated as fixed background. Occasionally, very bright nearby stellar
objects show significant residuals at the PSF core after subtraction; in these
cases a few pixel rectangular region is masked at the core before the final BL Lac fit.
The occasional column or pixel defect or cosmic ray are similarly masked.
Finally we mask the pixels within 1 FWHM in diameter of the BL Lac core. In
roughly 25\% of the OPTIC images this was required in any case due to non-linear ADU levels.
However, we found that sampling noise at 1 FWHM could dominate the fit statistics,
so a consistent exclusion of pixels at the core was used for all objects. For a few of the
brightest objects, we extended the core exclusion region to 1.5 FWHM, while for
a few of the faintest, we used 0.5 FWHM. The final results were not sensitive to
the precise value used.

	For each BL Lac model the background was determined from an annulus starting
at 7$^{\prime\prime}$ containing 2-3$\times$ the pixel count of the PSF model region; 
for a few poor seeing images and BL Lacs with very extended hosts we used a larger annulus.

\subsection{Noise Model \& $\chi^2$ Minimization}

For all multi-parameter fits, the parameters are determined by minimization of the quantity:

\begin{equation}
\chi^2 = \displaystyle\sum_{i} \frac{(I_{i} - M_{i})^2}{\sigma_i^2}
\end{equation}

Where the sum is over the $\{i\}$ unmasked pixels in the fitting region, $I_{i}$ is the 
observed intensity in ADU, $M_{i}$ is the model intensity in ADU
and $\sigma_i^2$ is the expected variance. 
The model is the sum of the constant background, PSF and host galaxy terms, where the 
analytic host model is integrated over a given pixel to determine the galaxy contribution.
The variance is estimated according to simple Poisson statistics to be:

\begin{equation}
\sigma_i^2 = \frac{I_i \times g + R^2}{g^2 \times (n_{im} - 1)}
\end{equation}

Where for OPTIC the gain was $g = 1.45 \ e^- {\rm ADU}^{-1}$ and for MiniMo 1.4$\ e^-
{\rm ADU}^{-1}$. The read noise $R$ is
$R = 4 \ e^-$ (OPTIC) or 5.5$\ e^-$ (MiniMo) and our final image stack is the from the median of $n_{im}$ 
normalized single exposures. This is a simplification since for OPTIC the orthogonal transfer
during integration correlated the sky noise (likely an artifact of imperfect flat-fielding). Also
the residual pixel noise in the unmasked region adds to the variance for sharp images. Nevertheless
we find that this variance is a reasonable first approximation to the local noise, and we can determine
the best-fit model by $\chi^2$ minimization. We use the simulated annealing algorithm from
Press et al. (1986).

	The final size of the simplex in $N+1$ dimensions gives the relative errors between the $N$
estimates of the fitting parameters. We would like to relate this to the 1$\sigma$ statistical
error, to provide a standard statistical error estimate for each parameter. We do this by scaling to
an externally estimated 1$\sigma$ error determined for each point source flux $\sigma_{PSF}$.
%\newpage

\subsection{Statistical Uncertainty}

	We determine this error by analogy with standard aperture photometry. In our case 
the central FWHM is masked in the fit so we compute the fractional error in the PSF amplitude 
for an annulus outside the masked region of width 1 FWHM as
\begin{equation}
f_{ann} = \frac{1}{N_{ann}} \times \sqrt{\frac{N_{ann}}{g \times (n_{im} - 1)} + n_{ann} \times \sigma_{sky}^2 + \frac{n_{ann}^2 \times \sigma_{sky}^2}{n_{sky}}}
\end{equation}
Where $N_{ann}$ is the source flux (ADU) within the photometry annulus, $n_{ann}$ is the number of 
pixels within the annulus, $\sigma_{sky}$ is the standard deviation of the background (ADU), 
and $n_{sky}$ is the number of pixels used to determine the background value. The first term under 
the radical is Poisson noise. The second is a statistical uncertainty due to the random 
fluctuations and unresolved sources in the background. The final term accounts for the 
possibility of a Poisson error in background level. The first term generally dominates, 
and thus $f_{ann}$ is $\sim \sqrt{2} \times f_{ap}$ as the annulus contains roughly 
1/2 the flux of an aperture without the central mask. 
We found a typical $f_{ann}$ for our PSF stars of $\sim 1-2 \times 10^{-3}$.

	With the large number of counts for the point sources measured here, even small
centroiding errors can in principle contribute to the error, since a varying fraction of the point source
falls in the central masked pixels. To test for this in each PSF fit, we artificially
shifted the model PSF position by $\pm 0.01$\,pixels, the maximum centroiding error, in each coordinate.
The rms variation in the included counts was computed to estimate $f_{cent}$, the fractional
error in flux estimation due to imprecise centroiding. Although this error only became
comparable to $f_{ann}$ for a few of the brightest point sources, we sum it in quadrature
to compute our statistical photometry error as 
$\sigma_{PSF} = N_{PSF}\sqrt{(f_{ann}^2 + f_{cent}^2)}$, where $N_{PSF}$ is the total
number of counts for the point source in ADU.
%Better check that Aaron actually did this???!!!

	As noted above, the size and shape of the simplex determined from
our $\chi^2$ minimization provides {\it relative} fitting errors. We adjusted
our convergence criterion such that the rms spread about the mean fit value for
the PSF amplitude was always close to $\sigma_{PSF}$ which, since the
nuclear point source always dominates the model, is a very good estimate
of the true model-fit statistical error.  Thus, by rescaling the
simplex amplitude such that the rms spread in the nucleus flux estimates
was equal to this photometric error $\sigma_{PSF}$,
we have normalized the simplex-determined error to be 1$\sigma$ (statistical) in each
quantity. We call this error $\sigma_{\rm stat}$.

\subsection{Systematic Uncertainty}

	Since our models are dominated by the bright core PSF, which is
different for each exposure, and since we have a limited number of stars
in each frame to generate a PSF model, our final uncertainty must be
dominated by systematic errors due to imperfections in this model. This
is in contrast to the HST-based fitting of \cite{urry00p1}, where
the PSF is, of course, stable and well-modeled and unresolved background
structure dominated the final uncertainty. We therefore
wish to estimate our systematic PSF uncertainty.

	We do so by fitting for a de Vaucouleurs host around unresolved stellar
sources. In each image stack we choose such test stars relatively close to the BL Lac
with comparable brightness. On average we were able to measure seven such stars per 
image. Ideally none would have been used in forming the PSF, but since these are the
stars most suitable for PSF modeling, they were largely included.
Before fitting, each star was treated exactly as its associated BL Lac, including
the same masking of central pixels. We then fit for the PSF amplitude and `host'
amplitude, with the de Vaucouleurs $R_{\rm e}=1.64^{\prime\prime}$ fixed at our default
radius. The fit host amplitude could either be negative (indicating 
excess wings on the model PSF) or positive (PSF model too narrow). Note that decreasing
the host angular size would weaken constraints on its amplitude, since it would
be more highly covariant with the PSF. However, our adopted value (appropriate
for a typical $R_{\rm e}=10$\,kpc at $z=0.5$) is reasonably conservative; even at the
minimum angular size at $z\sim1.6$, this would only decrease by 28\%.

\begin{figure}
\centering
\mbox{\subfigure{\includegraphics[width=3.5in]{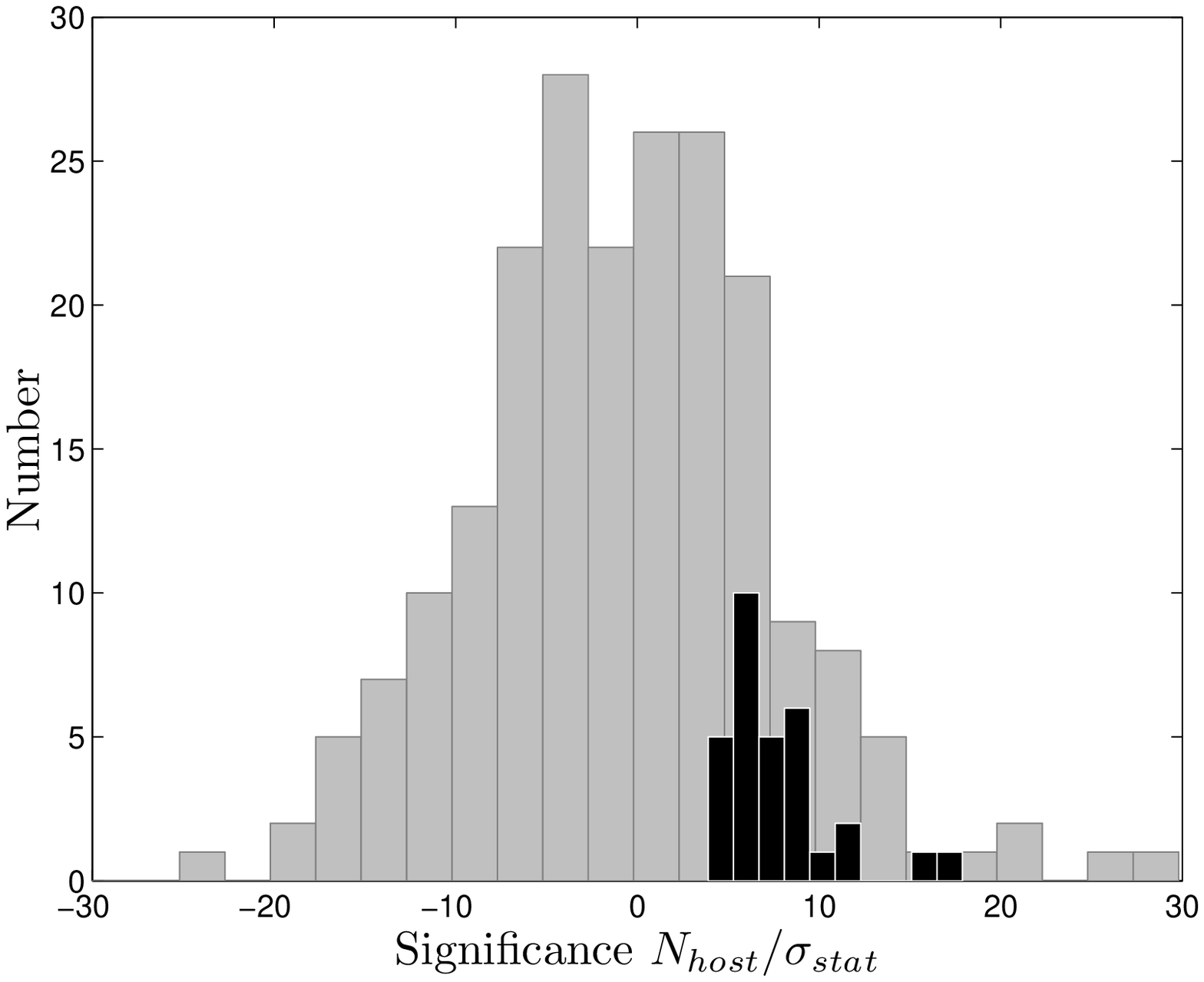}} %\quad\qquad
\subfigure{\includegraphics[width=3.5in]{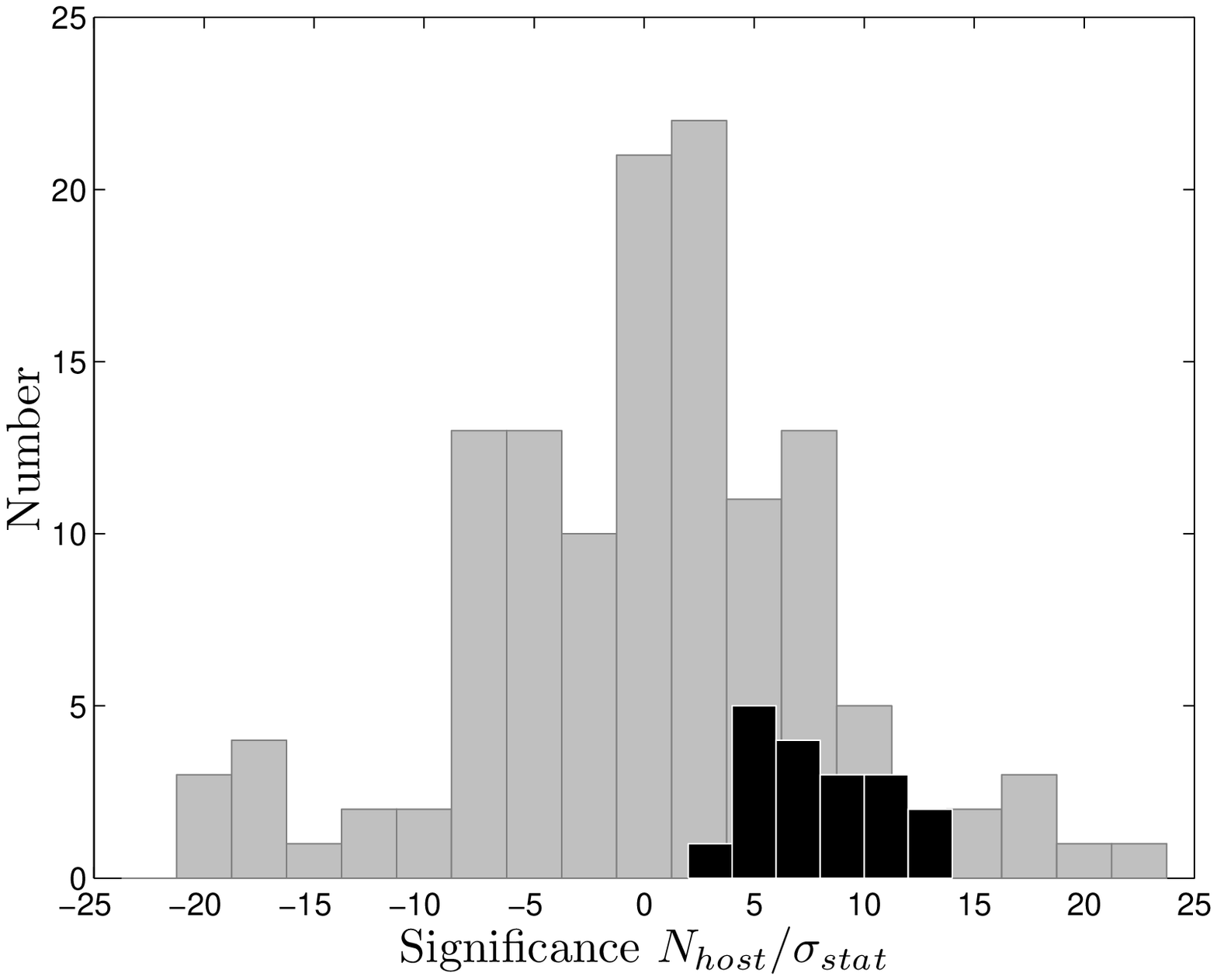} }}
\caption{Statistical significance of `host' detections around
field stars. Left: OPTIC stars for 31 fields, Right: MiniMo stars for
18 fields. The dark histograms show
the distribution of the rms significances for the individual BL Lac fields.
There is a modest tail, but most cluster near the global rms. These
individual rms values give our systematic error estimates for the individual
BL Lac fields.}
\label{fig:systematic}
\end{figure}

	The dispersion in the amplitudes of the fitted `host' fluxes provides
an estimate of our systematic uncertainty. However, since the stellar 
fluxes covered a substantial range, we elected to scale the `host' flux
significance to the statistical error on this flux for each test star $\sigma_{\rm stat}$
The rms spread in the `host' fluxes fit around the
test stars, in units of $\sigma_{\rm stat}$, then provides an estimate of the 
systematic error, for each individual BL Lac field. Figure \ref{fig:systematic} shows the
distribution of `host' significance ($N_{\rm host}/\sigma_{\rm stat}$) for the test 
stars in the OPTIC and MiniMo fields, along with the distribution of the rms 
significance for the individual fields.

	The rms significance of these `host' fits for OPTIC
and MiniMo were 8.3$\sigma_{stat}$ and 8.0$\sigma_{stat}$, respectively. This
suggests that our statistical (Poisson) errors on the host flux
substantially underestimate the true errors by $\sim 8\times$. 
However, it should be noted that the distributions are approximately normal and that
the mean fitted host flux is not significant. For the
OPTIC data, we obtain $-0.7 \sigma_{stat}$ while for MiniMo the average is
0.1$\sigma_{stat}$. In view of the systematic errors we expect random
offsets of $\pm 0.6\sigma_{stat}$ and $\pm 0.7\sigma_{stat}$ for the two
sets of test stars. We conclude that the frame-to-frame spread in
upper limits on surrounding host galaxies are appreciably larger than 
expected from pure Poisson statistics -- this can be attributed to errors in
forming the individual PSF models and to unresolved structure in the
PSF stars' and test stars' backgrounds. However there is no {\it overall}
introduced host flux; implying that our PSF model is a fair, albeit
uncertain, representation of the data.

	To be conservative, we adopt for our final error the {\it sum} of the
statistical and systematic errors, namely 
$\sigma_{fin}=\sigma_{stat}(1+\sigma_{syst})$, 
where $\sigma_{syst}=(N_{host}/\sigma_{stat})_{rms}$ is the rms value of the individual test stars in
the stacked image of an individual BL Lac. This is conservative since while
we expect the statistical errors to have a normal distribution, it would
be very surprising if the much larger systematic errors had wings as large
as a Gaussian out to many times the FWHM of the distribution. We tested this
error estimate by inserting Poisson realizations of model hosts around isolated
stars in several fields. For artificial galaxies assigned the total counts at 
the 3$\sigma_{fin}$ detection limit, we recovered host fluxes with rms errors of
$0.7 \sigma_{fin}$. In no case was the difference between the injected and recovered
counts larger than $1 \sigma_{fin}$.

\subsection{Final BL Lac Fitting Procedure}

	We constrain the properties of the BL Lac host by the hierarchical fit.	
We first fit for a circularly symmetric host with fixed angular radius 
$R_{\rm e} = 1.64^{\prime\prime}$, just as for the test star measurements.
The final uncertainty on the host counts is taken to be $\sigma_{fin}$ where
the statistical error is taken from the simplex rescaled to the aperture photometry
error for the BL Lac core, while the systematic error is the multiple of this
statistical error determined by `host' fits to the test stars in this BL Lac's
field, as described above.
If the host flux is larger than $3 \sigma_{fin}$ (which is dominated by
the systematic uncertainty), we deem it significant. If the significance does not reach
$3 \sigma_{fin}$, we infer an upper limit of $N_{host} + 3\sigma_{fin}$,
or $3\sigma_{fin}$, whichever is larger.

	If there is a significant detection of a BL Lac host, we next re-run
the fits allowing the host angular radius $R_{\rm e}$ to vary. The only
fitting constraint is $R_{\rm e} > 0$. If the best fit value of $R_{\rm e}$ is less than the 
FWHM for the image, we do not deem it significant. A value for $R_{\rm e}$ is
only quoted if $R_{\rm e} > 3 \sigma_{fin}$, where the final statistical plus
systematic error is estimated from the re-scaled simplex, as above.
Significant estimates of $R_{\rm e}$ were found for 25\% of the OPTIC-measured BL Lacs
and 33\% of those measured with MiniMo. Note that the overall flux and its fit error
were free to vary, as well. As a consistency check, we note that in no case where a
significant $R_{\rm e}$ measurement was found, did the host flux measurement decrease
in significance to below $3\sigma$.

	Finally, we attempted to fit for a significant host ellipticity,
with two additional parameters $\epsilon$ and the position angle $\theta$ added to
the model. These parameters both exceeded $3\sigma_{fin}$ significance for
only two of the BL Lacs observed with OPTIC. Again the host flux and size
$R_{\rm e}$ were free in this final fit.

	Tables \ref{optic_results} and \ref{minimo_results} contain the 
final fit nucleus and host amplitudes (converted to magnitudes, see below) 
and the host physical size (when available). % and their combined $\sigma_{fin}$
%(statistical+ systematic) uncertainties for the BL Lacs observed during
%the two campaigns. 
Details of the observations (exposure lengths, FWHM
of the stellar sources and the number of stars used in the PSF model) 
are also listed.

\section{Redshift Estimates \& Lower Limits}

\begin{figure}
\begin{minipage}[b]{0.5\linewidth} % A minipage that covers half the page
\centering
\includegraphics[width=9.5cm]{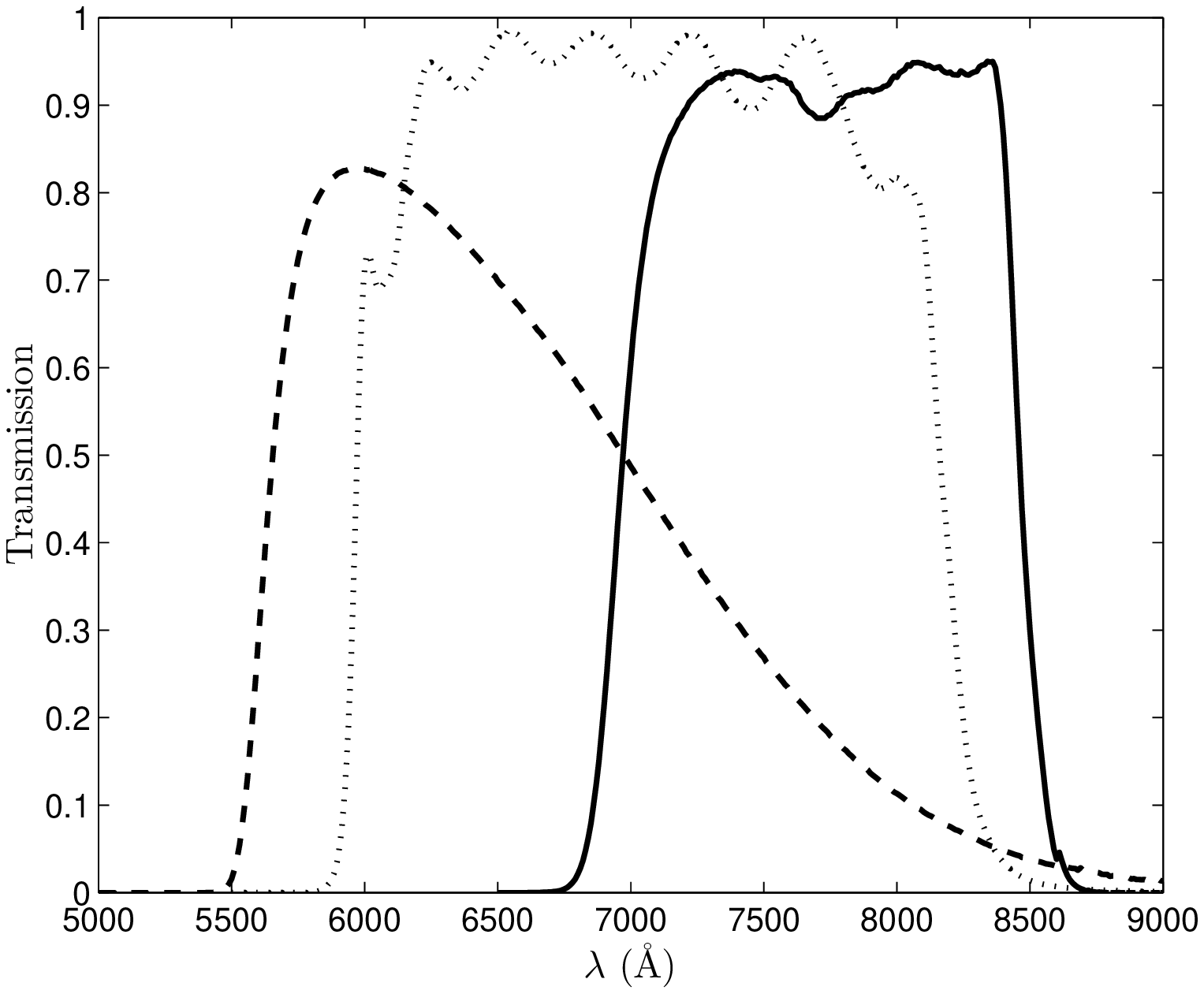}
\caption{\label{fig:filters}  Filter transmission curves used in this study.
Solid $i^\prime$, dashed $R$, dotted {\it HST} F702W. We convert from
the assumed host $z=0$ $R$ magnitude to our observed $i^\prime$ magnitude.
Note that the $i^\prime$ cut-on is redder than used in previous studies.}
\end{minipage}
\hspace{-0.1cm} % To get a little bit of space between the figures
\begin{minipage}[b]{0.5\linewidth} % A minipage that covers half the page
\centering
\includegraphics[width=9.5cm]{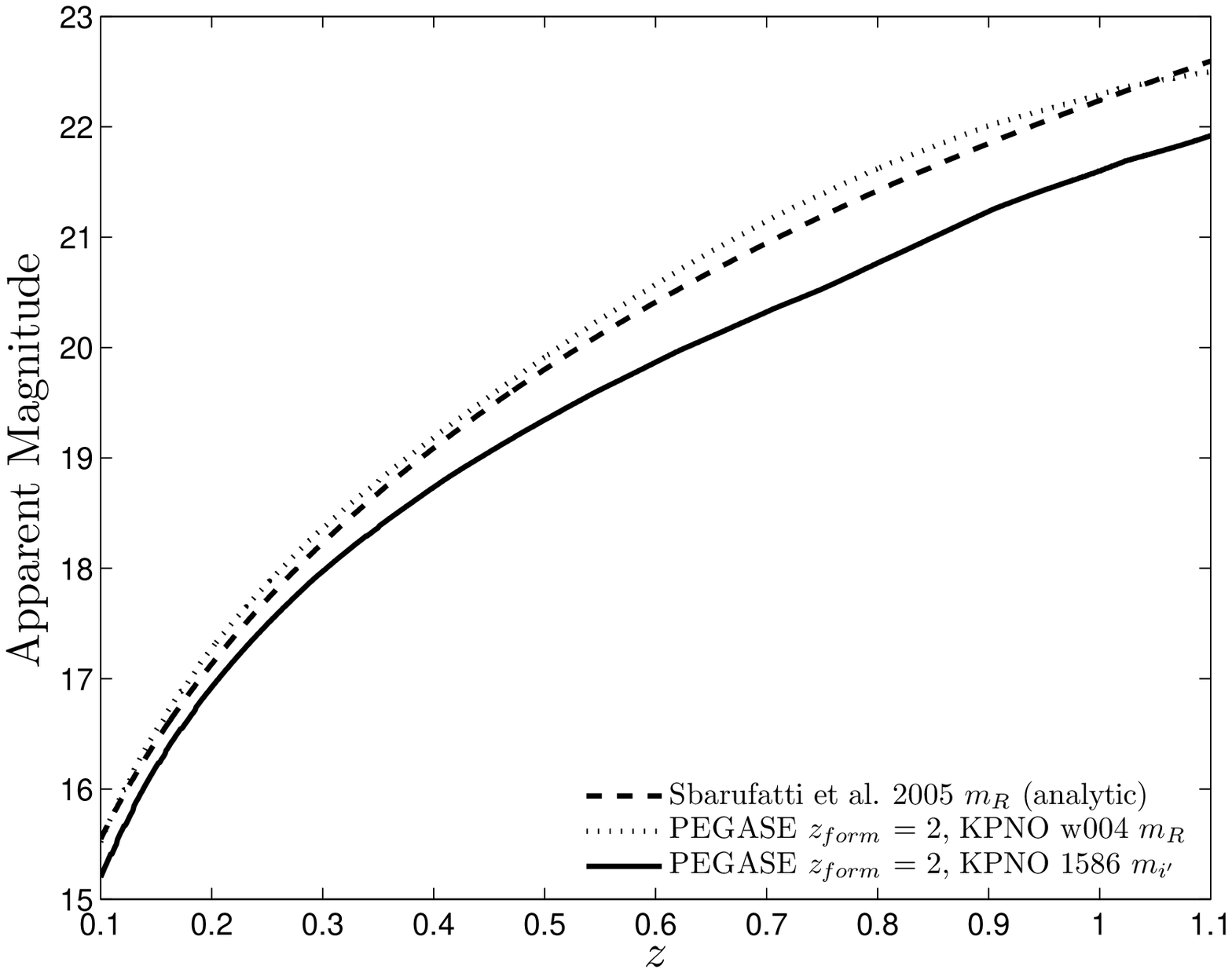}
\caption{\label{fig:hubble_diagram_all} Hubble diagrams calculated in KPNO w004 $R$ and
KPNO 1586 $i^\prime$. The analytic estimate for the $R$ Hubble diagram of \cite{sba05} 
is shown for comparison.}
\end{minipage}
\end{figure}

	Since \citet{sba05} have argued that BL Lac host galaxies are
standard candles with $M_R = -22.9 \pm 0.5$, we can use our measurements and
upper limits on the host flux to extract redshift estimates and lower bounds.
To do this we compute an $i^\prime$ Hubble diagram for our KPNO 1586
filter. We improve on the similar R-band Hubble diagram of \citet{sba05},
by including host evolution, assuming a host (elliptical) formation redshift
of $z_{form} = 2$ \citep{urry05}. With this assumption, we adopt an 
elliptical galaxy spectrum computed at the appropriate age from the PEGASE
model of \citet{frv01}, where the age is computed for each $z_{obs}<z_{form}$
for our standard cosmology. To lock the overall normalization to
the $M_R$ of \citet{sba05}, we fixed the flux of the $z=0$, evolved elliptical
spectrum to $M_R = -22.9$. This required folding through an $R$ filter.
Unfortunately the precise KPNO Kron-Cousins $R$ filter used is not recorded,
except in a private 1995 communication between J. Holtzman and Landolt \citep{hol95}.
Accordingly we use the transmission curve for Kron-Cousins R filter KPNO w004,
first obtained in 1994; this may be the precise filter used in the original study.
In any case, after normalizing with this filter at $z=0$, we take the model
flux from each observation redshift, convert to the observed frame using
the luminosity distance $d_L$ from our standard cosmology and convolve with the 
$i^\prime$  KPNO 1586 transmission (Figure \ref{fig:filters}) to obtain 
the apparent magnitude.

	The resulting Hubble diagram is shown in Figure \ref{fig:hubble_diagram_all}.
For comparison we have also computed the $R$-band Hubble diagram for this evolving
model. The curve is quite close to \citet{sba05}'s analytic approximation, although
departures caused by the evolution are evident. Hubble diagram computations 
with other evolving galaxy models yielded similar curves, with differences small
compared to those introduced by uncertainty in the host absolute magnitude.

	To compare with our measurements, we converted our model fit host
counts (in ADU) to $i^\prime$ magnitudes, including atmospheric \citep{massey02}
and Galactic \citep{schlegel98} extinction corrections. The establishment of the flux 
scale is discussed above (\S3), while the assumed Galactic extinctions from the dust 
maps at the BL Lac locations are listed in 
Tables \ref{optic_results} and \ref{minimo_results}.

\begin{figure} [h]
\begin{center} 
\scalebox{.6}{\includegraphics{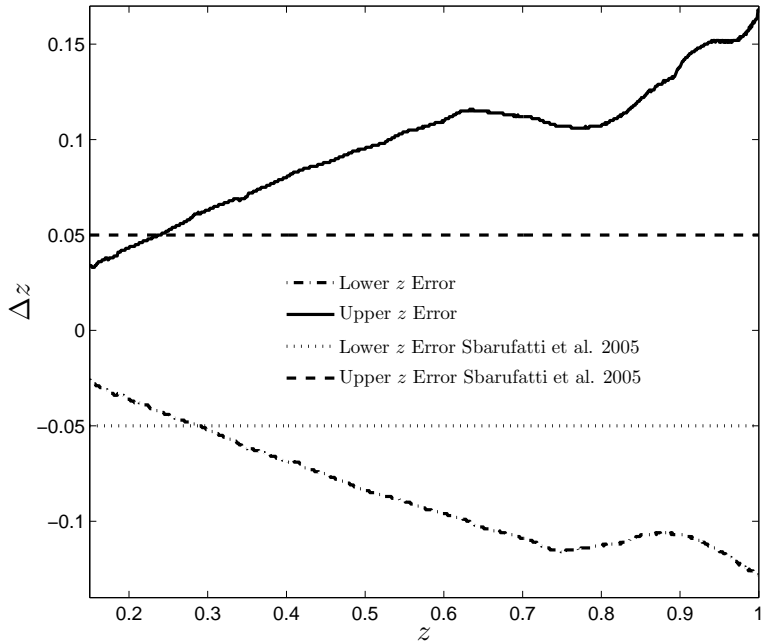}}
\caption{\label{fig:error_comparison} Upper and lower error error bars as 
a function of redshift, inferred from the $\pm 0.5$ dispersion in
host absolute magnitude $M_R$ at $z=0$.}
\end{center}
\end{figure}

	Finally, these host magnitudes (and lower limits) are converted to
redshifts using our Hubble diagram. Of course the host absolute magnitude uncertainty
translates into a range of possible redshifts. There are two contributions to
the full redshift uncertainty. The first is host measurement uncertainty, including
both statistical and systematic photometry errors.  We use the $\pm \sigma_{fin}$
range for the maximum and minimum host magnitude to infer asymmetric error
bars on the host redshifts. This is augmented by adding, in quadrature, the
photometry errors estimated for our zero point -- this increase is however very small.
These values are listed as the first error flags in the final columns 
of Tables \ref{optic_results} and \ref{minimo_results}.
In addition, the claimed $\delta R= 0.5$\,mag dispersion in the
host luminosity leads to a redshift range. We propagate the $z=0$ $R$ absolute 
magnitude uncertainty through observed $i^\prime$ and hence to $\Delta z$ as
a function of redshift (see Figure \ref{fig:error_comparison}).
This nearly always completely dominates
the redshift uncertainty and is listed as the second error flag in our tables.

	For objects with a fit angular $R_{\rm e}$, the central redshift estimate is used to 
convert to proper kpc; the values are comfortably close to the otherwise assumed
$R_{\rm e}=10$\,kpc. Finally, for undetected hosts, we 
take the minimum $z$ allowed by the $3\sigma_{fin}$ host magnitude limit, 
the photometric scale errors, and the standard value for
the host luminosity to obtain a lower limit on the host redshift. Again this 
is listed in the last column of 
Tables \ref{optic_results} and \ref{minimo_results}.

	We have treated our measurement errors conservatively, having
summed the systematic and statistical uncertainties in our error flags
and requiring $3\times \sigma_{fin}$ detections before claiming a
measurement is significant.  On the other hand, we consider only the assumed 
dispersion in the host luminosity. Clearly an improved calibration 
of the host magnitudes (if uniform) can make these measurements much 
more useful.

\section{Results and Discussion}

\begin{figure} [h]
\begin{center}
\scalebox{.5}{\includegraphics{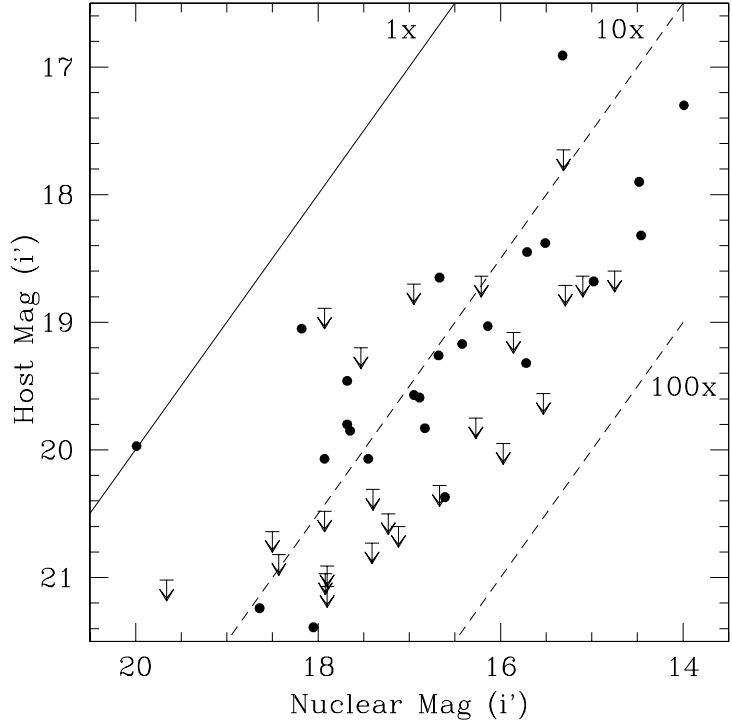}}
\end{center}
\caption{\label{fig:NucHost} Host magnitude $vs.$ nucleus magnitude.
The dashed lines show hosts $10\times$ and $100\times$ fainter than the
BL Lac nucleus.}
\end{figure}

	We were able to detect hosts for 14/31 OPTIC-observed BL Lacs 
and 11 of 18 observed with MiniMo. The redshift estimates varied from 
$\sim 0.20 - 0.95$ and the lower limits from $\sim 0.25 - 0.9$. For OPTIC
the median value was $z=0.54$ for MiniMo $z=0.53$. It is interesting to
compare this with the {\it HST}\, host detections of \citet{urry00p1}; in this study
nearly all BL Lacs at $z<0.5$ were resolved but only 6/23 sources with known
$z>0.5$ yielded host detections. While at first sight it might seem surprising
that a ground-based program could be competitive, it should be remembered that 
the {\it HST} snapshot BL Lac survey observed with the F606W and F702W filters. 
The $i^\prime$ filter blue cut-on is $\sim 100$\,nm redder than for
the bulk of the {\it HST} observations, and so we do not expect to suffer
appreciable surface brightness attenuation from the 4000\AA\,  break until
$z \sim 0.75$. Further the typical HST snapshot exposures were $\sim 500$\,s
with a 2.4\,m aperture $vs.$ 900$-$2100\,s with a 3.6\,m aperture. The combination
of redder band and deeper exposure allow some additional sensitivity to the low surface
brightness host wings. Of course, our sources were selected from those
{\it lacking} spectroscopic redshifts, making it likely that these are a higher
redshift sub-sample. However it is also worth noting that the typical
$\sim 10-30\times$ core host ratio plotted in Figure \ref{fig:NucHost}
represents an even stronger core dominance, compared to the {\it HST}\,
sample, than one might think
since the continua of BL Lacs is quite blue compared to the hosts,
and our $i^{\prime}$ bandpass is redder than that of previous $R$ or
$F606W/F702W$ studies.

For 14 of the 25 detected hosts we obtained estimates of the host angular size. The inferred 
median effective radii were 9.2 kpc (OPTIC) and 8.0 kpc (MiniMo) in good agreement
with previous BL Lac host estimates \citep{urry00p1,urry00p2,urry00p3}. The other detected
hosts had $R_{\rm e}$ measurements with $<3\sigma_{fin}$ significance or best fit radii
smaller than the FWHM and thus were judged to be insignificantly resolved.

\begin{figure}
\centering
\mbox{\subfigure{\includegraphics[width=3in]{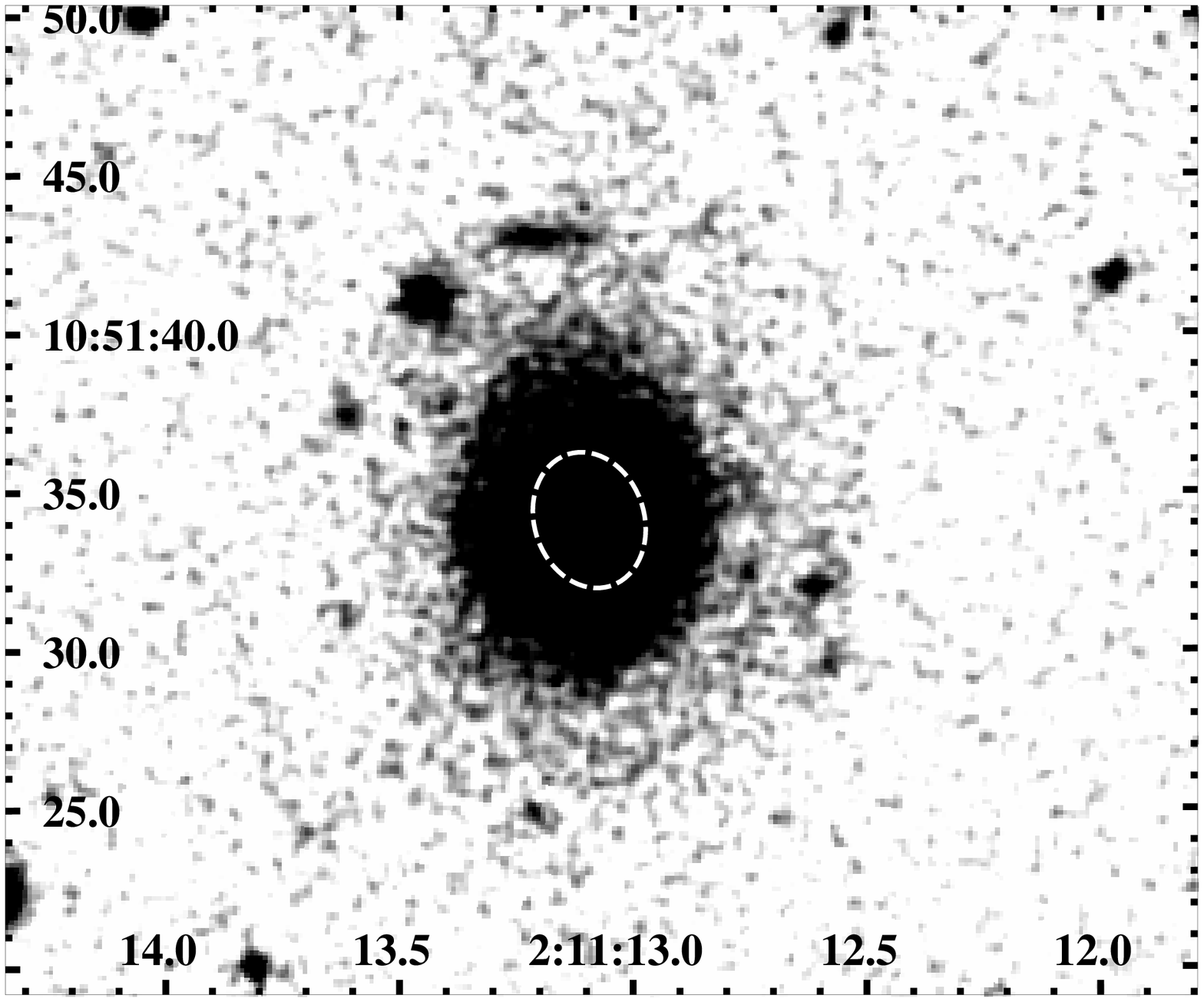}}\quad
\subfigure{\includegraphics[width=3in]{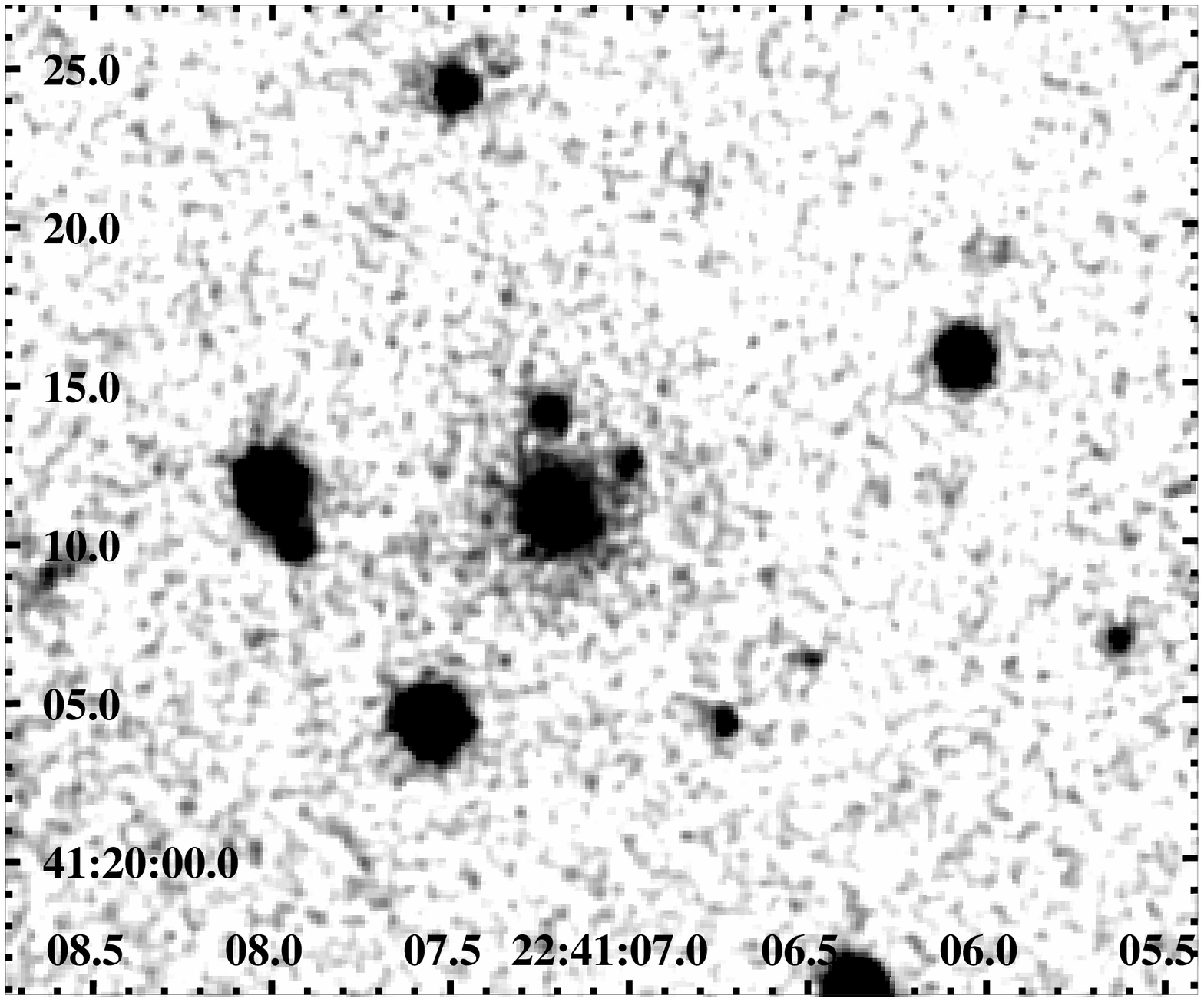} }}
\caption{Two BL Lac fields after core PSF subtraction. Left:
J0211+1051 at $z\approx 0.20$, showing the highly resolved host and several
fainter companions. The host has significant ellipticity; 
the fit $R_{\rm e}$ ellipse is shown.
Right: J2241+4120 at $z\approx 0.52$, showing the faint 
wings of the host and two fainter companions. No significant 
ellipticity is detected.
} \label{fig:coresub}
\end{figure}

In two cases we have a formal detection of ellipticity. For 
J0211+1051 the small value $\epsilon=0.21$ accords well with the visual impression
(Figure \ref{fig:coresub}). In the other case, J0348$-$1610, the ellipticity is
also small $\epsilon=0.14$, but here the result is suspect, as there is a
bright companion within $2 R_{\rm e}$. Indeed it may be no coincidence that
the two objects with significant $\epsilon$ both have a large number of companions.
While the upper limits on $\epsilon$ are in some cases constraining,
we conclude that with our modest ground-based resolution we cannot in most
cases probe the host structure.

	One arena where the comparison with HST-based measurements should be fairly
robust is the host-nucleus flux ratio. Here we find a marked difference with the HST
sample.  Of the 69 BL Lac hosts resolved by \citet{urry00p1}, 54\% of these had 
$f_{host}/f_{nucleus} \gtrsim 1$. Even counting the 42 unresolved BL Lacs 34\% had
a nuclear-host ratio $\gtrsim 1$. In contrast, we find only 1 object of our 49 BL Lacs 
has $f_{host}/f_{nucleus} > 1$. Indeed
63\% of our hosts are $\ge 10\times$ fainter than the nucleus (Figure \ref{fig:NucHost}).
Two selection effects may account for this difference. First, these objects have been
selected as bright flat spectrum radio core sources. Further a significant fraction
of the MiniMo targets were in addition known to be active gamma-ray emitters.
This implies that the Earth line-of-sight is even more nuclear component dominated (on-axis)
than for the typical BL Lac. Second, these are the subset of these systems {\it lacking}
previous spectroscopic redshifts. Again, this suggests that they should be even more
core continuum dominated than the typical BL Lac.

\subsection{Radial Profile Plots}
	
\begin{figure}
\centering
\mbox{\subfigure{\includegraphics[width=3in]{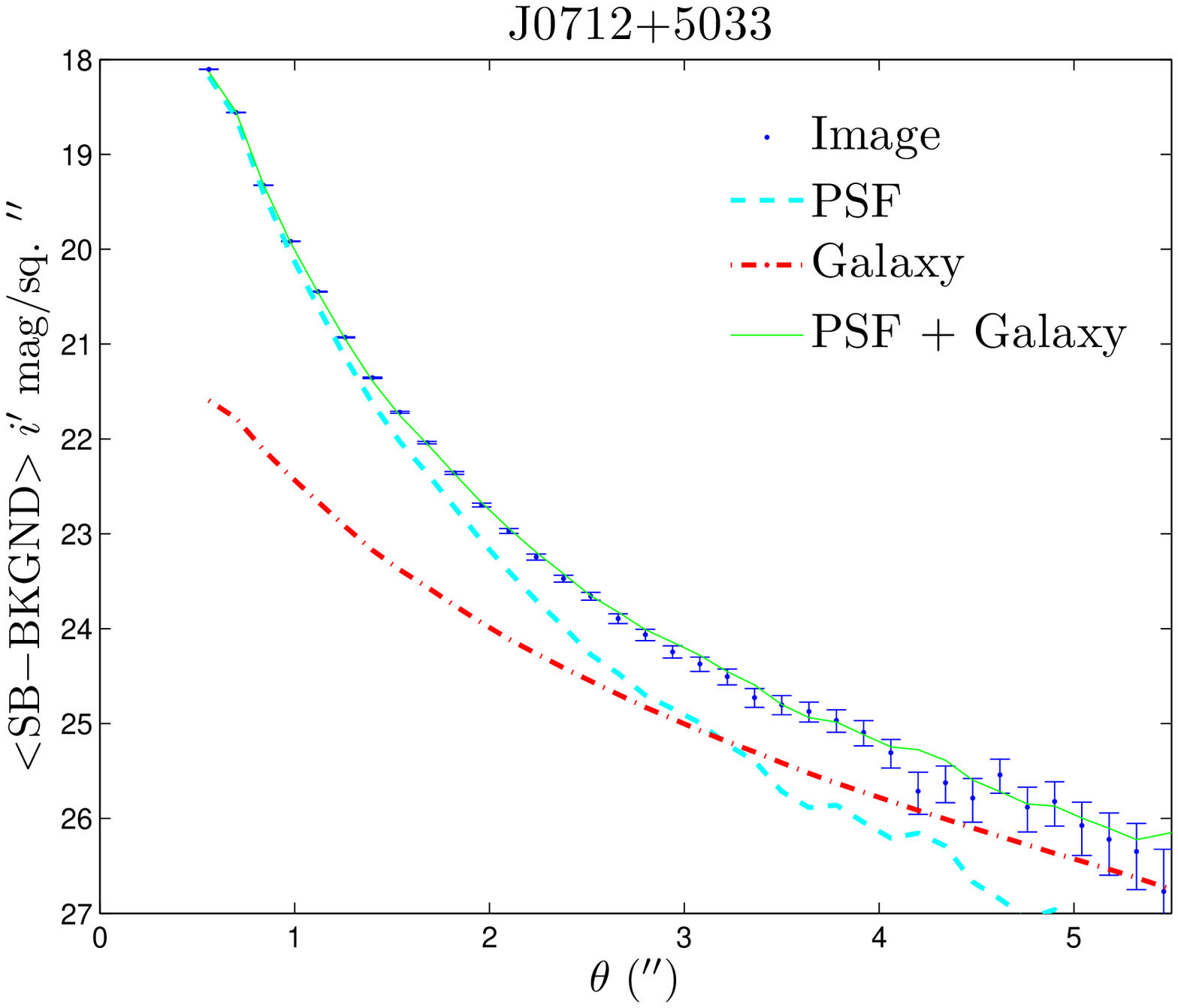}}\quad
\subfigure{\includegraphics[width=3in]{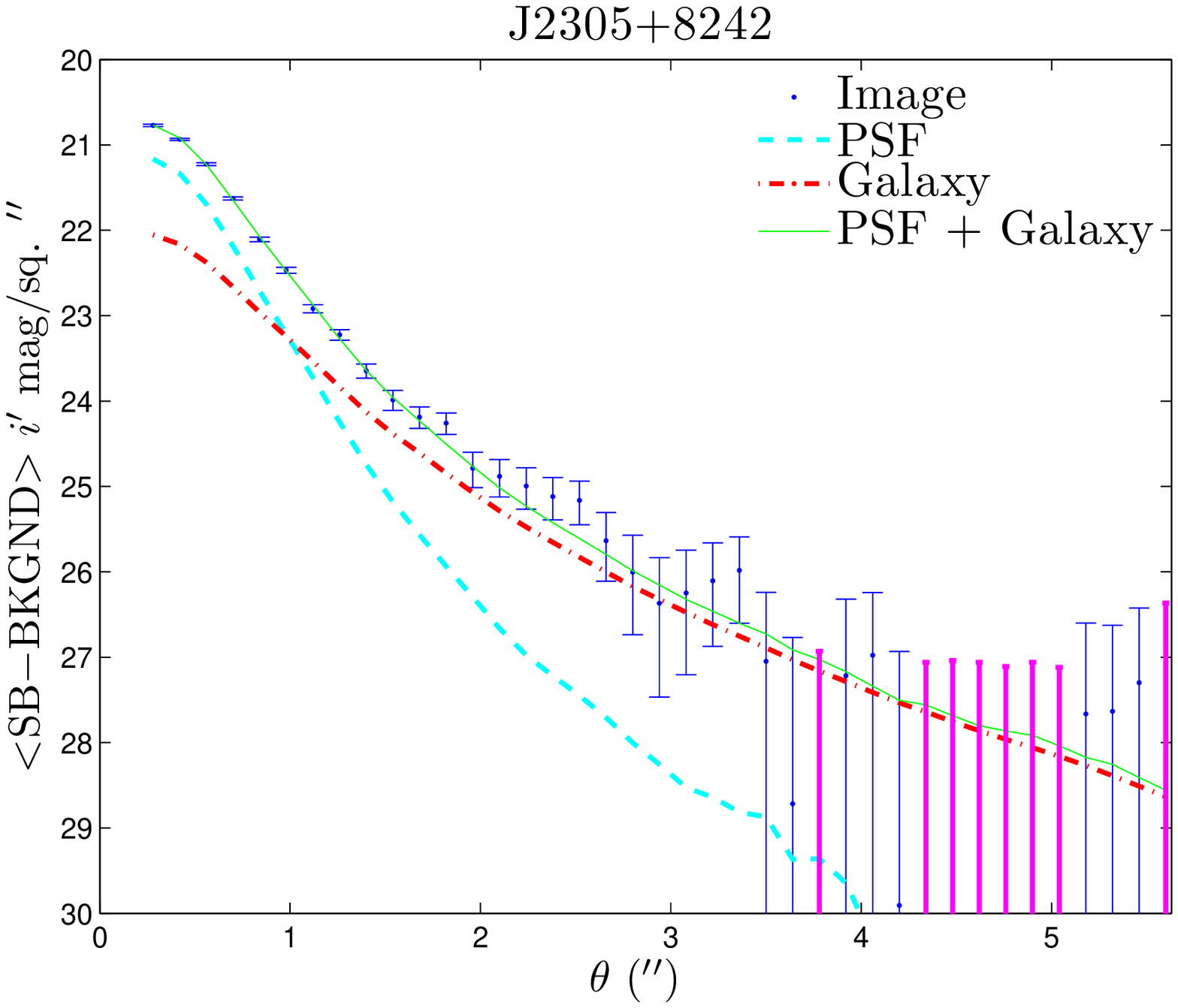} }}
\caption{OPTIC radial profiles for fit BL Lacs: Left: J0712+5033, resolved at $z\approx0.47$.
Right: J2305+8242, resolved at $z\approx0.62$.
The dashed lines show the PSF, the dot-dashed lines the host and
the solid line the total model. Error flags on individual radial bins
are statistical only.
} \label{fig:radOPdet}
\end{figure}

\begin{figure}
\centering
\mbox{\subfigure{\includegraphics[width=3in]{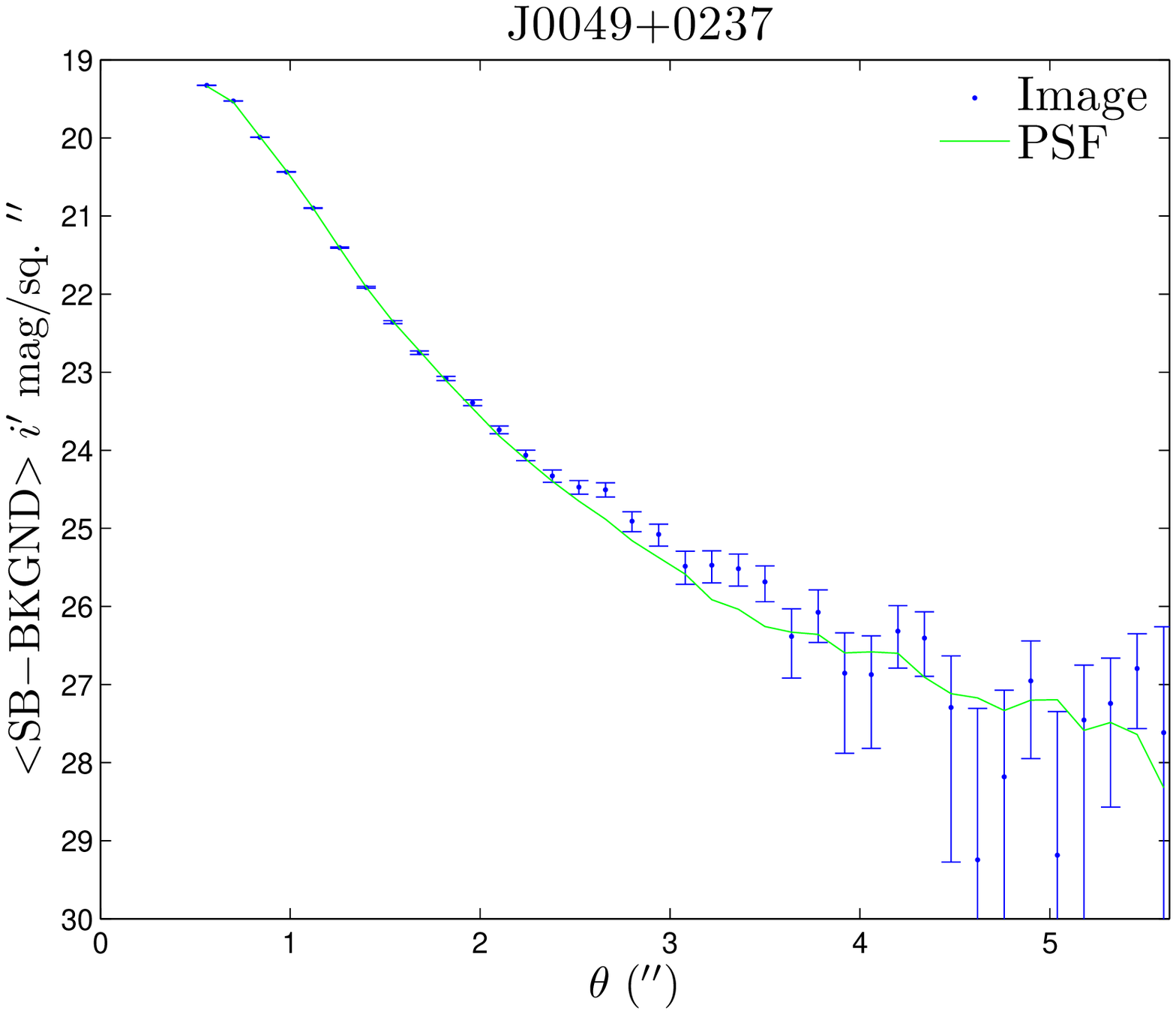}}\quad
\subfigure{\includegraphics[width=3in]{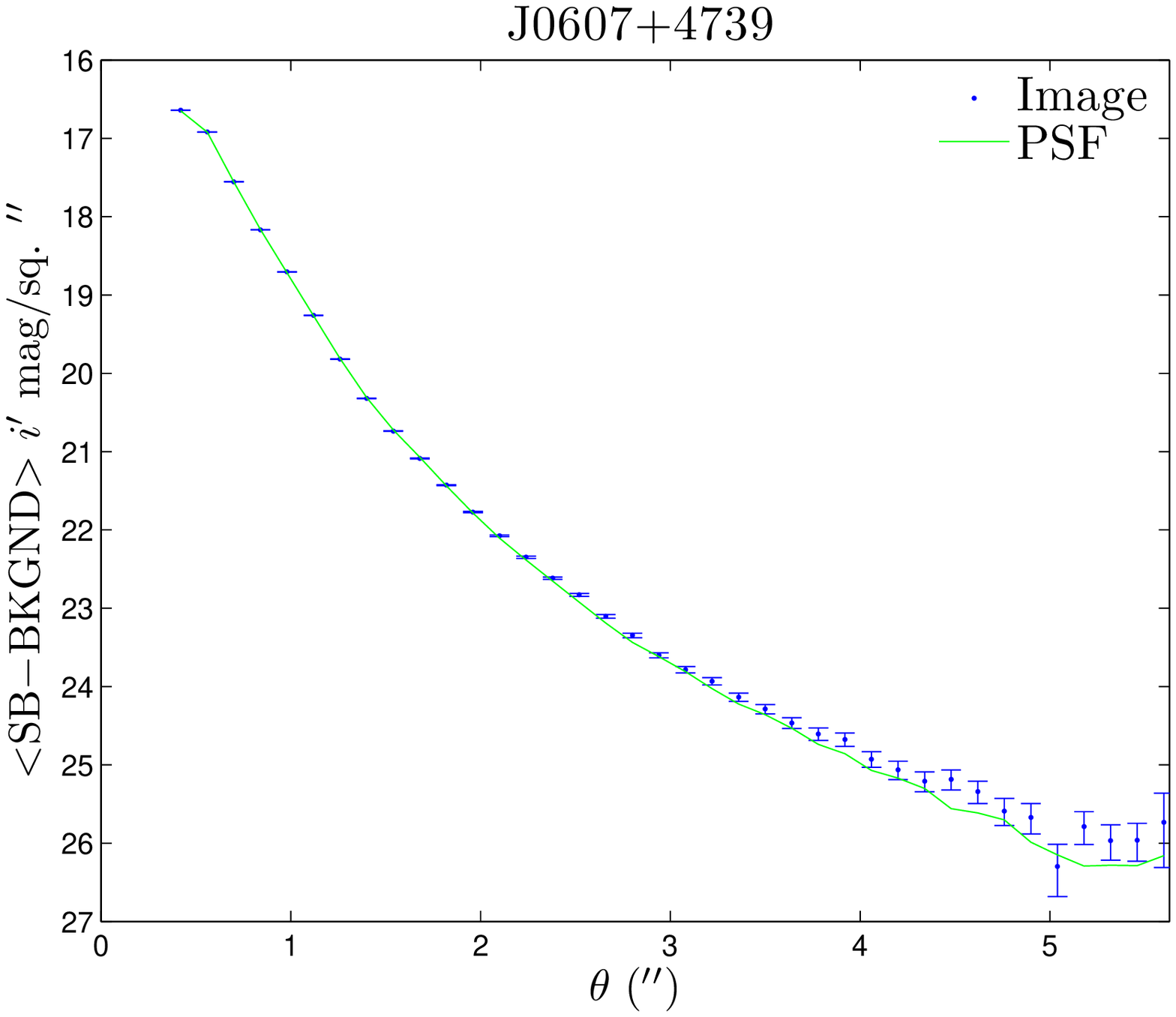} }}
\caption{OPTIC radial profiles for fit BL Lacs with host non-detections: 
Left: J0049+0237, unresolved $z>0.84$. Right: J0607+4739, unresolved $z>0.54$.
} \label{fig:radOPul}
\end{figure}

\begin{figure}
\centering
\mbox{\subfigure{\includegraphics[width=3in]{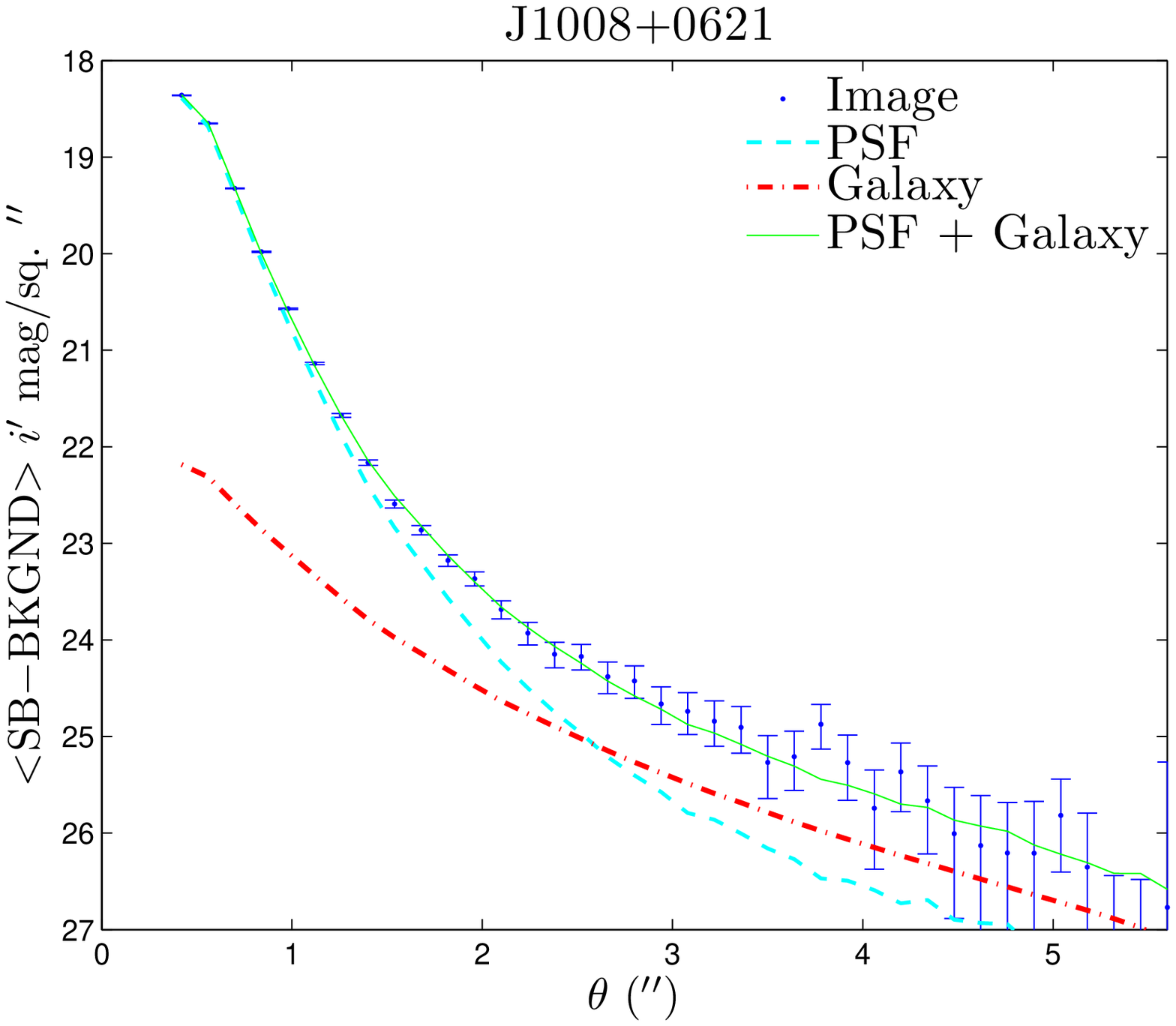}}\quad
\subfigure{\includegraphics[width=3in]{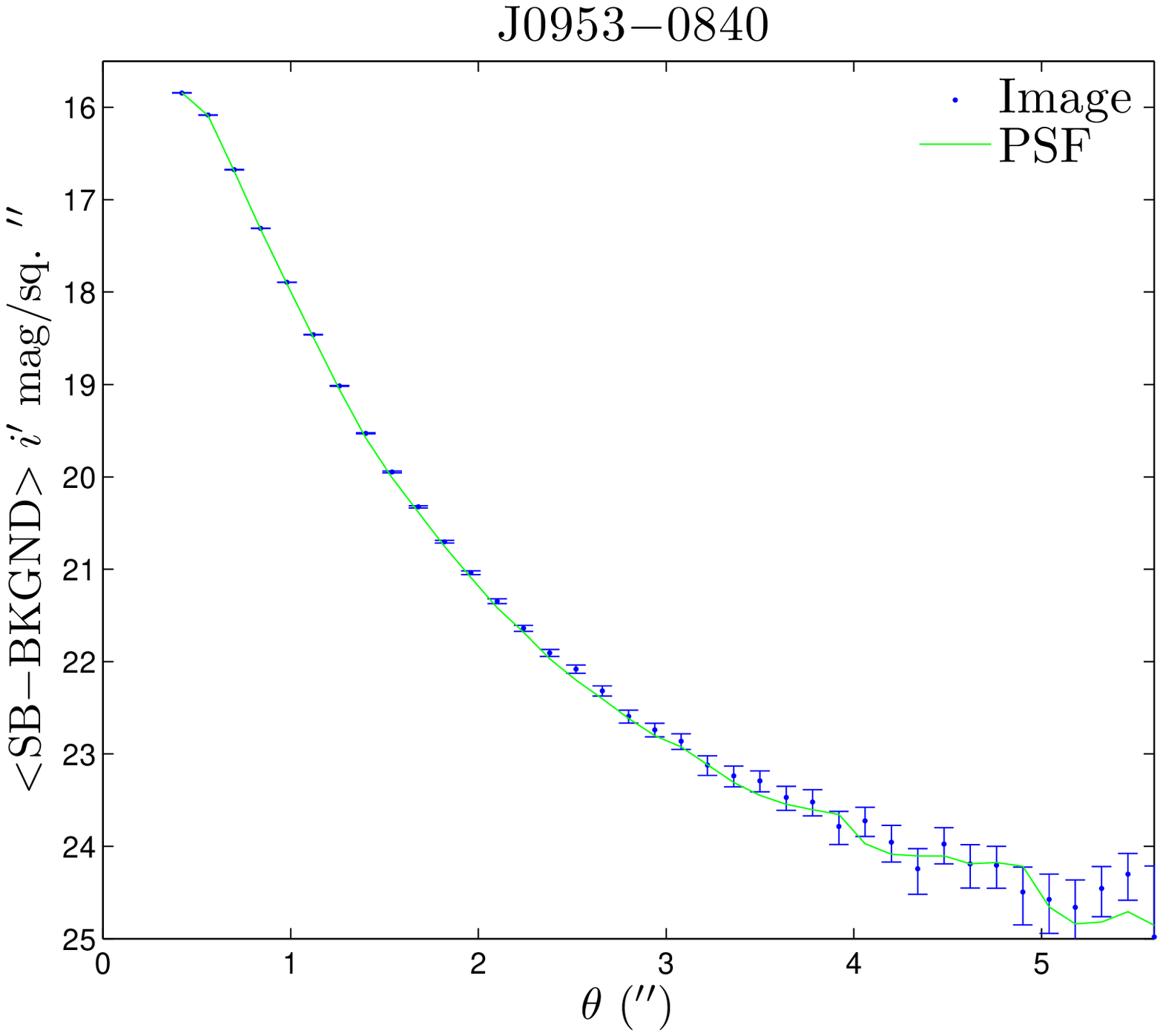} }}
\caption{MiniMo radial profiles for fit BL Lacs.
Left J1008+0621, resolved at $z\approx 0.59$.
Right: J0953$-$0840, unresolved $z>0.40$.
} \label{fig:radMi}
\end{figure}

	We present here a sample of azimuthally averaged $i^\prime$ instrumental 
surface brightness plots, compared with our best fit model components
(Figures \ref{fig:radOPdet} $-$ \ref{fig:radMi}). The surface brightness 
data are measured from excess counts above the above the fitted 
background level measured in annuli about the core position. The model curves
are from integrations over these radial bins.  The error bars shown are 
1$\sigma$, calculated in a fashion consistent with the variance in \S5. 
%Which Sigma???/
We show two BL Lacs with host detections using the OPTIC camera, as well as
two fits providing upper limits. For MiniMo we show one detected host and
one unresolved source.

\subsection{Near Environments}

	It has been noted \citep{urry00p1,urry00p2,urry00p3} that BL Lac hosts are
surrounded by a significant excess of nearby galactic objects. Following
\citet{urry05}, we count companions within a projected proper distance of 50\,kpc
of the BL Lac centroid. When we do not have a $z$ estimate, we count only
companions within $5.9^{\prime\prime}$ of the nucleus (50\,kpc at the minimum
angular diameter distance $z \sim 1.6$). We count only resolved objects.

We find at least one companion for 51\% of our BL Lacs (25/49), in reasonable agreement
with the detection rate of \citet{urry00p2} (47\%) and \citet{urry05} (62.5\%). 
Of course, with our ground-based imaging, some of the unresolved nearby sources
may represent additional compact galactic companions. The number of companions $N_c$
within a projected distance of $r_\perp = 50$\, kpc, the magnitudes of the two brightest
companions (and the comparison of the brightest to the host flux $\Delta m_{1,host}$)  
and a flag
indicating evidence for interaction are given in Table \ref{companionship}. As noted
above, our two hosts with nominal detections of ellipticity have both large
numbers of nearby companions and morphological evidence for interaction.
These may in fact represent recent mergers in compact groups. 
J1440+0610 has an extended companion with centroid at $r_\perp > 50$\,kpc which extends
within 50\,kpc. For two objects (J0050$-$0929 and J0743+1714) 
the companion at small radius may be physically contiguous with the undetected host.
Clear evidence for interaction
was seen only for the better-seeing OPTIC data, so additional companion
interaction is likely.

\begin{table} [h]
\begin{center}
\begin{tabular}{c c c c c c c c}
Name & $N_{c}$ & $m_{i^{\prime},1}$ & $m_{i^{\prime},2}$ & $\Delta m_{1,host}$ & $r_{min}$ ($''$) & $r_{min}$ (kpc) & Int.? \\
                        \hline \\ [-2ex]
                        J0050$-$0929 & 1 & $>$ 19.5 & - & - & $<$ 0.6 & - & M \\
                        J0202+4205 & 1 & 22.7 & - & 1.3 & 4.1 & 33 & N \\
                        J0203+7232 & 1 & 22.3 & - & - & 5.6 & - & N \\
                        J0211+1051 & 7 & 21.9 & 23.6 & 5.0 & 7.3$\dagger$ & 24$\dagger$ & Y \\
                        J0219$-$1842 & 1 & 21.5 & - & 1.7 & 3.6 & 25 & Y \\
                        J0348$-$1610 & 5 & 21.5 & 21.6 & 2.9 & 2.7 & 14 & Y \\
                        J0607+4739 & 1 & 23.0 & - & - & 4.6 & - & Y \\
                        J0610$-$1847 & 1 & 23.3 & - & 3.2 & 4.9 & 35 & N \\
                        J0625+4440 & 2 & 22.5 & 24.2 & - & 3.8$\dagger$ & - & M \\
                        J0650+2502 & 2 & 21.6 & 22.5 & 2.6 & 5.3$\dagger$ & 31$\dagger$ & N \\
                        J0743+1714 & 2 & $>$ 21.8 & 23.7 & - & $<$ 0.6 & - & M \\
                        J0814+6431 & 1 & 22.3 & - & 3.9 & 5.2 & 25 & M \\
                        J0817$-$0933 & 2 & 22.4 & 23.8 & 2.0 & 4.9$\dagger$ & 35$\dagger$ & N \\
                        J0835+0937 & 1 & 22.2 & - & - & 3.8 & - & M \\
                        J0907$-$2026 & 1 & 21.9 & - & 3.6 & 8.0 & 39 & N \\
                        J0915+2933 & 1 & 22.1 & - & 3.7 & 9.4 & 47 & N \\ 
                        J1008+0621 & 3 & 22.1 & 23.7 & 2.3 & 4.1 & 27 & M \\
                        J1243+3627 & 1 & 23.9 & - & 4.6 & 4.1 & 25 & M \\
                        J1253+5301 & 1 & 24.1 & - & 4.3 & 7.5 & 50 & M \\
                        J1427+2347 & 1 & 22.7 & - & 5.4 & 8.1 & 30 & M \\
                        J1440+0610 & 1 & 20.2 & - & 0.6 & 8.1 & 64 & N \\
                        J1542+6129 & 1 & 22.8 & - & 4.1 & 7.7 & 40 & M \\
                        J1624+5652 & 1 & 23.0 & - & 2.1 & 5.0 & 35 & N \\
                        J2241+4120 & 2 & 22.6 & 23.0 & 3.5 & 2.8$\dagger$ & 18$\dagger$ & Y \\
                        J2305+8242 & 1 & 23.1 & - & 3.1 & 5.6 & 42 & N \\
                        \hline
 \end{tabular}
 \end{center}
 \caption{\label{companionship} List of brightest observed companions within $r_\perp = 50$\,kpc. 
Except for those marked $\dagger$, the brightest companion is also the closest.
We classify the morphological evidence for interaction as obvious ``Y," 
possible ``M" or absent ``N."}
\end{table}

%\newpage
\bigskip

\subsection{Comparison with Other Redshift Estimates}

	As a part of our on-going spectroscopic campaign to obtain 
identifications and redshifts for the {\it Fermi}-detected blazars,
some of these imaging targets have had additional spectroscopic exposure.
In a few cases we obtain true spectroscopic redshifts $z_s$. In several
other cases we have lower limits on the redshift $z_{abs}$ from detection
of intervening intergalactic absorption-line systems. Finally, in several
cases we were able to use spectroscopic limits on absorption line-strengths
(typically Ca H \& K and G-band limits), together with the assumption of 
a uniform host magnitude (as in this paper) to place lower redshift
limits $z_{HK}$ on the BL Lac \citep{shaw09}.
\smallskip

	We have obtained a direct redshift measurement for only one of these
targets, using Keck LRIS spectroscopy \citep{shaw10}. The value is within 
$1\sigma$ (statistical+systematic) of the imaging estimate:
\smallskip

\qquad J2022+7611 $z_I=0.49^{+0.04+0.09}_{-0.03-0.08}$ ($z_s=0.584$). 
\smallskip

\noindent
For three objects our imaging $z_I$ estimates are in agreement with lower limits
from our own spectroscopy \citep{shaw09}. The low redshift of 
J1427+2347 is particularly interesting in view of the recent VERITAS
TeV detection of this source (Ong et al. 2009).
\smallskip

\qquad J0712+5033 $z_I=0.47^{+0.03+0.09}_{-0.02-0.08}$ ($z_{HK}>0.47$).	%Shaw et al.

\qquad J1427+2347 $z_I=0.23^{+0.01+0.05}_{-0.01-0.04}$($z_{HK}>0.03$). 

\qquad J1440+0610 $z_I=0.55^{+0.02+0.10}_{-0.02-0.09}$($z_{abs}>0.316$).	

\smallskip\noindent
For two objects our imaging redshift constraints are stronger than those
obtained from the spectroscopic constraint on the HK line strengths:
\smallskip

\qquad J0909+0200 $z_I>0.83$ ($z_{HK}>0.54$, \citet{shaw09}.	%shaw et al.

\qquad J0049+0237 $z_I>0.84$ ($z_{HK}>0.82$ \citet{sba06}.  %sbarufatti et al.

\smallskip\noindent
For two objects our lower limit on the redshift from imaging is not
as strong as that from our spectroscopy \citep{shaw09}:	
\smallskip

\qquad J0050$-$0929 $z_I>0.27$ ($z_{HK}>0.44$) and		%Shaw et al.

\qquad J2050+0407 $z_I>0.74$ ($z_{abs}>0.819$).

\smallskip\noindent
Finally for two objects the imaging redshifts are in disagreement with
the lower limits from limits on the HK absorption line strength:
\smallskip

\qquad J1253+5301 $z_I=0.59^{+0.02+0.11}_{-0.02-0.10}$ ($z_{HK}>0.77$; $1.4\sigma$ higher) and %shaw eta l.

\qquad J1542+6129 $z_I=0.39^{+0.03+0.08}_{-0.03-0.07}$ ($z_{HK}>0.63$; $2.2\sigma$ higher).	%Shaw et al.

\smallskip
\smallskip
	These last two disagreements may partly be attributed to
differences in the host size and slit losses assumed in the spectroscopic
method. However,
statistically we should expect some $z_I$ estimates to be low,
as Malmquist bias will assure that fluctuations causing detections
make the sources appear artificially close. Of course it is also likely
that outliers in the BL Lac host magnitude distribution exist: these
will cause larger disagreements. As we continue to collect spectroscopic
redshifts and limits for these sources, we should be able to probe the
fraction of sub-luminous hosts and test the utility of the standard candle
hypothesis of \citet{sba05}.

\section{Conclusions}

\begin{figure} [h]
\begin{center}
\scalebox{.5}{\includegraphics{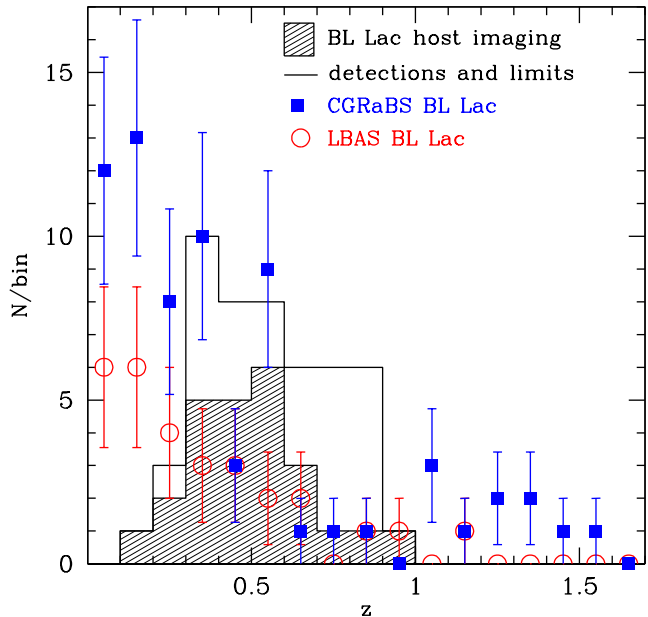}}
\end{center}
\caption{\label{fig:zhist} BL Lac redshift distributions. 
Estimates from host detections (filled histogram), and with
host lower limits (line histogram). For comparison we also
show the spectroscopic BL Lac redshift distribution from the two parent
populations providing our target lists: Squares -- the CGRaBS
flat spectrum radio loud BL Lacs \citep{het08} and
Circles -- the early {\it Fermi} gamma-ray-detected BL Lacs \citep{ab09a}.
}
\end{figure}

	We have detected hosts for half of the BL Lac objects imaged
in this WIYN campaign. These detections, assuming a standard host
luminosity, give redshift estimates with a median value of 
$z_{\rm med}=0.51$. The upper limits on host flux for the remaining objects
also give reasonable redshift constraints with a median value
for the bound of $z_{\rm med}>0.61$. This may be compared with the 
spectroscopic redshifts already obtained for the two parent BL
Lac populations: for the BL Lacs in the early {\it Fermi}
blazar list one finds $z_{\rm med} = 0.33$, while for the
radio selected CGRaBS BL Lacs, we have $z_{\rm med} = 0.45$ (Figure \ref{fig:zhist}).
We can further quantify the difference by making Kolmogorov-Smirnoff comparisons
of the distributions. While the LBAS and CGRaBS sets are
quite consistent, with a KS probability of 0.58, the imaging
set differs from both with a KS probability of similarity of
only $1-5 \times 10^{-3}$. If we include the lower $z$
bounds in the distribution the probability drops to
$<2 \times 10^{-5}$.

	The evident lack of low redshift imaging detections
is doubtless a selection effect: such objects will in general
show the strongest absorption features from the host and thus
are most likely to provide spectroscopic redshifts. With the 
ground-based imaging results, we obtain a number of additional
redshift estimates, with good efficiency for host detection out to $z\sim 0.65$. The
distribution of our upper limits suggests that space-based imaging
searches, using near-IR filters, can be productive to even
higher redshift. In any event, it appears that it would be a mistake
to assume that the objects in the radio and gamma-ray
samples lacking spectroscopic redshifts are similar to
those with spectroscopic distances. The imaging results
suggest, in contrast, that these other sources are appreciably
more distant and, on average have higher luminosity. It
will be important to include these imaging $z$ estimates
and bounds in attempts to measure the BL Lac luminosity function.

	One other major conclusion of this analysis is that
host searches should be at the $i$ band or redder. Here the
contrast with the relatively blue nuclear continuum flux 
is improved: we have seen that this allows host detections
in sources with extreme core dominance.  The increased
sensitivity to the low surface brightness halos of the redshifted
hosts should also materially improve the detection fraction at
$z > 0.5$. Again this will be important for study of the
population and evolution of BL Lac hosts galaxies. Finally,
it will be important to pursue spectroscopic confirmations 
of these redshift estimates, as only with such data will it
be possible to test and extend the claimed uniformity of
the BL Lac host population and to probe its possible evolution.

\acknowledgements

This work was supported in part by NASA Grant NNX08AW30G through the Fermi Guest 
Investigator Program.
We thank the staff of the WIYN observatory for excellent support and in particular Steve Howell
for help with OPTIC and Di Harmer for the MiniMo set-up. We in addition thank J. Windschitl
for sharing notes on OPTIC reduction techniques and Mike Shaw and Steve Healey for
help with comparison to the spectroscopic BL Lac sample.

\appendix

	Here we note peculiarities of the fields of individual BL Lacs,
particularly those that required amendment to the standard analysis
described above.
\medskip

\noindent
{\bf J0050$-$0929}:
This high Galactic latitude field is particularly sparse, with a deficit of suitable of PSF stars (13). 
The BL Lac host shows asymmetric structure to E, which we flag in Table \ref{companionship} 
as due to a close companion.
Even with a relatively good 0.59$^{\prime\prime}$ FWHM its nature is not clear. In fact,
the fixed $R_{\rm e}$ fit suggests a resolved host with significance $\sim 2.5\sigma_{fin}$,
but gives an exceptionally high $\chi^2/\textrm{D.O.F.} \sim 12$. This implies that the source
is not well modeled by a de Vaucouleurs host. Thus we interpret the $3\sigma$ non-detection
as a redshift lower limit, inferring a distance consistent with the \citep{shaw09} spectroscopically
derived bound $z > 0.44$.
\smallskip

\noindent
{\bf J0110+6805}:
This target is at $|b|<10^\circ$ and so is not a member of the CRATES catalog. It
is otherwise known as 4C+67.04.
An unresolved object of comparable brightness appears  $\sim 2$ FWHM from
the AGN nucleus. We did a PSF subtraction of this star and masked the
peak pixels. Since our PSF fits are sequential, not simultaneous, the
substantial overlap with the BL Lac core PSF introduced additional uncertainty
in the BL Lac host fit.  We approximate this by adding the aperture
photometry PSF uncertainties of the neighbor star and BL Lac nucleus
in quadrature, using this as the nucleus statistical uncertainty and propagating through
the rest of the analysis.

\smallskip\noindent
{\bf J0203+7232}:
A bright $i \sim 14$ star appears $\sim 4''$ from the BL Lac centroid.
As for J0110+6805 we increase the nucleus photometric error by adding 
in the uncertainty in the amplitude fit for this star.

\smallskip\noindent
{\bf J0211+1051}:
The detectable BL Lac host extends beyond the standard fitting radius $5.6^{\prime\prime}$;
accordingly we extend the radius of the fitting region to $11.2^{\prime\prime}$,
and increase the size of the sky annulus.

\smallskip\noindent
{\bf J0348$-$1610}:
A very bright $i \sim 11$ field star $20^{\prime\prime}$ from the BL Lac
produced noticeable scattering wings across the fitting region. We estimated
this scattered flux by radially averaging the stellar wings over regions free of
background objects and then subtracted this flux from the fitting region
to remove the small induced gradient. After subtraction, we applied the
standard BL Lac fitting.

\smallskip\noindent
{\bf J0743+1714}:
Two bright ($i \sim 14.5$)  stars are $\sim 20^{\prime\prime}$ and $25^{\prime\prime}$
from the BL Lac. Subtraction of a radially averaged template of the scattering wings,
as for J0348$-$1610, produced a flat background acceptable for the BL Lac fitting.

\smallskip\noindent
{\bf J0814+6431}:
This BL Lac extends beyond the standard fitting region. This is increased
to $7.7^{\prime\prime}$; the sky annulus radius is increased accordingly.

\smallskip\noindent
{\bf J0817$-$0933}:
Two bright ($i\sim 14$)  stars are $\sim 25^{\prime\prime}$ and $30^{\prime\prime}$
from the BL Lac. Subtraction of a radially averaged template of the scattering wings,
as for J0348$-$1610, produced a flat background acceptable for the BL Lac fitting.

\smallskip\noindent
{\bf J0907$-$2026}:
With a bright nucleus and relatively poor seeing, we started the
background sky annulus at $10.5^{\prime\prime}$. The BL Lac fitting
still employed the standard $5.6^{\prime\prime}$ radius.

\smallskip\noindent
{\bf J0915+2933}:
The scattering wing of a bright $i\sim 13$ star $\sim 50^{\prime\prime}$
away was removed, as for J0348$-$1610.

\smallskip\noindent
{\bf J1253+5301}:
The scattering wing of a bright $i\sim 13.5$ star $\sim 40^{\prime\prime}$ 
away was removed, as for J0348$-$1610.

\smallskip\noindent
{\bf J1427+2347}:
The bright BL Lac nucleus is highly saturated, resulting in bleeding. In this 
case the central exclusion was increased to diameter of 3 FWHM and the bleed 
trail was carefully masked. The test stars were fit with a matching 
central exclusion region. While the sky background annulus radius was
increased to $16.8^{\prime\prime}$, the BL Lac model fit was performed in
the standard $5.6^{\prime\prime}$ radius region.

\begin{sidewaystable}
  \begin{center}
    \begin{tabular} {c c c c c c c c c c c}
      Name & Exposures (s) & Seeing ($''$) & $b \ (^{\circ})$ & A$_{I}$& N$_{PSF \ stars}$ & $i^\prime_{nucleus}$ & $i^\prime_{host}$ & $f_{host}/f_{nucleus}$ & $R_{\rm e}$ (kpc) & $z$ \\ [.2ex]
      \hline \\ [-2ex]
      J0004$-$1148 & [180,180,300,300,300] & 0.74 & $-71.1$ & 0.06 & 25 & 19.66 & $>$ 21.02 & $<$ 0.28 & - & $>$ 0.86 \\ [.3ex]
      J0049+0237 & [300,300,300,300,300] & 1.04 & $-60.3$ & 0.05 & 29 & 17.92 & $>$ 20.97 & $<$ 0.06 & - & $>$ 0.84 \\ [.3ex]
      J0050$-$0929 & [60,60,60,60,60,60,60,60,60] & 0.59 & $-72.4$ & 0.06 & 13 & 15.31 & $>$ 17.65 & $<$ 0.12 & - & $>$ 0.27 \\ [.3ex]
      J0110+6805$\tablenotemark{a}$ & [180,180] & 0.76 & 5.29 & 2.38 & 61 & 14.48 & 17.90$_{-.28}^{+.36}$ & 0.04 & - & 0.29$_{-.03-.05}^{+.03+.06}$ \\ [.3ex]
      J0202+4205 & [180,300,300,300,300] & 0.46 & $-18.9$ & 0.18 & 24 & 18.05 & 21.39$_{-.31}^{+.42}$ & 0.05 & $13.9 \pm 1.58$ & 0.94$_{-.07-.11}^{+.13+.15}$\\ [.3ex]
      J0203+7232 & [180,180,180] & 0.49 & 10.4 & 1.37 & 52 & 16.21 & $>$ 18.64 & $<$ 0.11 & - & $>$ 0.39 \\ [.3ex]
      J0211+1051 & [60,60,60,180,180,180] & 0.50 & $-47.4$ & 0.27 & 27 & 15.32 & 16.91$_{-.09}^{+.09}$ & 0.23 & $6.34 \pm 0.17$ & 0.20$_{-.01-.04}^{+.01+.04}$ \\ [.3ex]
      J0219$-$1842 & [180,180,180,180,180] & 0.76 & $-68.1$ & 0.07 & 20 & 17.65 & 19.85$_{-.16}^{+.18}$ & 0.14 & - & 0.60$_{-.03-.10}^{+.04+.11}$ \\
      J0348$-$1610 & [180,180,180,180,180] & 0.70 & $-47.6$ & 0.09 & 24 & 16.67 & 18.65$_{-.07}^{+.07}$ & 0.16 & $11.1 \pm 2.33$ & 0.39$_{-.01-.07}^{+.01+.08}$ \\ [.3ex]
      J0433+2905 & [180,180,300,300,300] & 0.46 & $-12.6$ & 1.49 & 30 & 17.53 & $>$ 19.20 & $<$ 0.21 & - & $>$ 0.48 \\ [.3ex]
      J0502+1338 & [300,300,300,300,300] & 0.49 & $-16.8$ & 0.99 & 28 & 18.18 & 19.05$_{-.08}^{+.08}$ & 0.44 & $8.87 \pm 1.73$ & 0.45$_{-.01-.08}^{+.01+.09}$ \\ [.3ex]
      J0509+0541 & [60,60,60,60,60] & 0.53 & $-19.6$ & 0.21 & 28 & 14.75 & $>$ 18.60 & $<$ 0.03 & - & $>$ 0.38 \\ [.3ex]
      J0527+0331 & [300,300,300,300,300,300,300] & 0.67 & $-16.9$ & 0.31 & 27 & 18.64 & 21.24$_{-.17}^{+.19}$ & 0.10 & $4.91 \pm 1.63$ & 0.90$_{-.04-.11}^{+.05+.14}$ \\ [.3ex]
      J0607+4739 & [180,180,180,180,180,180] & 0.78 & 12.9 & 0.36 & 30 & 15.53 & $>$ 19.56 & $<$ 0.02 & - & $>$ 0.54 \\ [.3ex]
      J0610$-$1847 & [180,180,180] & 0.83 & $-17.3$ & 0.17 & 24 & 17.45 & 20.07$_{-.31}^{+.42}$ & 0.09 & - & 0.64$_{-.06-.10}^{+.10+.12}$ \\ [.3ex]
      J0612+4122 & [130,90,120,120,120] & 0.50 & $10.9$ & 0.39 & 28 & 16.67 & $>$ 20.28 & $<$ 0.04 & - & $>$ 0.69 \\ [.3ex]
      J0625+4440 & [180,180,180,180,180] & 0.60 & 14.4 & 0.29 & 23 & 17.12 & $>$ 20.60 & $<$ 0.04 & - & $>$ 0.77 \\ [.3ex]
      J0712+5033 & [240,120,120,120] & 0.60 & 23.9 & 0.13 & 28 & 16.42 & 19.17$_{-.14}^{+.16}$ & 0.08 & $8.68 \pm 1.83$ & 0.47$_{-.02-.08}^{+.03+.09}$ \\ [.3ex]
      J0814+6431 & [90,90,90,90,90] & 0.45 & 33.2 & 0.12 & 26 & 15.51 & 18.38$_{-.09}^{+.10}$ & 0.07 & $11.4 \pm 3.06$ & 0.35$_{-.01-.06}^{+.01+.07}$ \\ [.3ex]
      J0909+0200 & [180,180,180,180,180] & 0.52 & 31.4 & 0.06 & 28 & 17.90 & $>$ 20.91 & $< 0.06$ & - & $>$ 0.83 \\ [.3ex]
      J1813+0615 & [180,180,180] & 0.83 & 11.3 & 0.41 & 106 & 17.40 & $>$ 20.31 & $<$ 0.07 & - & $>$ 0.70 \\ [.3ex]
      J190311+554044 & [120,120,120,120] & 0.53 & 20.5 & 0.12 & 36 & 16.27 & $>$ 19.75 & $<$ 0.04 & - & $>$ 0.58 \\ [.3ex]
      J1927+6117 & [120,120,120,120,120] & 0.69 & 19.5 & 0.13 &  70 & 16.95 & 19.57$_{-.10}^{+.11}$ & 0.09 & - & 0.54$_{-.02-.09}^{+.02+.10}$ \\ [.3ex]
      J2009+7229 & [180,180,180,180,180] & 0.66 & 20.2 & 0.53 & 30 & 17.23 & $>$ 20.50 & $<$ 0.05 & - & $>$ 0.74 \\ [.3ex]
      J2022+7611 & [180,180] & 0.64 & 21.1 & 0.47 & 37 & 16.68 & 19.26$_{-.18}^{+.21}$ & 0.09 & - & 0.49$_{-.03-.08}^{+.04+.09}$ \\ [.3ex]
      J2050+0407 & [180,180,180,180,180] & 0.83 & $-24.0$ & 0.18 & 29 & 17.93 & $>$ 20.48 & $<$ 0.10 & - & $>$ 0.74 \\ [.3ex]
      J2200+2137 & [300,300,300,300,300] & 0.67 & $-26.0$ & 0.20 & 32 & 18.50 & $>$ 20.64 & $<$ 0.14 & - & $>$ 0.77 \\ [.3ex]
      J2241+4120 & [300,300,300,300,300] & 0.70 & $-15.2$ & 0.47 & 31 & 17.68 & 19.46$_{-.11}^{+.12}$ & 0.19 & $9.60 \pm 1.00$ & 0.52$_{-.02-.09}^{+.02+.10}$ \\ [.3ex]
      J224356+202101 & [90,60,60] & 0.56 & $-33.4$ & 0.09 & 39 & 15.10 & $>$ 18.64 & $<$ 0.04 & - & $>$ 0.39 \\ [.3ex]
      J2305+8242 & [300,300,300] & 0.94 & $20.6$ & 0.46 & 31 & 19.99 & 19.97$_{-.15}^{+.15}$  & 1.03 & - & 0.62$_{-.03-.10}^{+.04+.12}$ \\ [.3ex]
      J2346+8007 & [300,300,300,300] & 1.01 & $17.6$ & 0.45 & 68 & 17.90 & $>$ 21.07  & $<$ 0.14 & - & $>$ 0.87 \\ [.7ex]
      \hline
    \end{tabular}
\tablenotetext{a}{J0110+6805=4C+67.04}
    \caption{\label{optic_results} Summary of OPTIC results. 4+4 names are from CGRaBS \citep{het08}, 6+6 names
are from CRATES \citep{het07}. There are also two  $>3\sigma$ dectections of host
ellipticity: J0211+1051 with $\epsilon = 0.21 \pm 0.02$ at $\theta =19 ^{\circ} \pm 10^{\circ}$ and 
J0348$-$1610 with $\epsilon = 0.14 \pm 0.04$ at $\theta =32 ^{\circ} \pm 15^{\circ}$ with angles
measured N through E.}
  \end{center}
\end{sidewaystable}
\clearpage

\newpage

\begin{sidewaystable}
  \begin{center}
    \begin{tabular} {c c c c c c c c c c c}
      Name & Exposures (s) & Seeing ($''$) & $b \ (^{\circ})$ & A$_{I}$ & N$_{PSF \ stars}$ & $i^\prime_{nucleus} $ & $i^\prime_{host}$ & $f_{host}/f_{nucleus}$ & $R_{\rm e}$ (kpc) & $z$ \\ [.2ex]
      \hline \\ [-2ex]
      J024330+712012 & [300,300] & 1.09 & 10.4 & 1.50 & 51 & 16.95 & $>$ 18.70 & $<$ 0.20 & - & $>$ 0.40 \\ [.3ex]
      J065047+250304 & [60,300,300,300,300] & 0.68 & 11.0 & 0.19 & 31 & 16.14 & 19.03$_{-.10}^{+.11}$ & 0.07 & - & $0.45_{-.02-.08}^{+.02+.09}$ \\ [.3ex]      
      J0743+1714 & [300,300,300] & 0.65 & 19.0 & 0.07 & 41 & 18.43 & $>$ 20.82 & $<$ 0.11 & - & $>$ 0.81 \\ [.3ex]
      J0817$-$0933 & [300,300,300,300] & 0.78 & 14.3 & 0.20 & 42 & 16.61 & 20.37$_{-.18}^{+.21}$ & 0.03 & $12.5 \pm 0.58$ & 0.71$_{-.04-.11}^{+.05+.11}$ \\ [.3ex]
      J083543+093659 & [300,300,300] & 0.83 & 27.5 & 0.10 & 30 & 17.93 & $>$ 18.89 & $<$ 0.41 & - & $>$ 0.42 \\ [.3ex]
      J0907$-$2026 & [300,300,300] & 0.97 & 18.0 & 0.34 & 35 & 14.46 & 18.32$_{-.20}^{+.25}$ & 0.03 & $8.05 \pm 0.39$ & $0.34_{-.03-.06}^{+.03+.07}$ \\ [.3ex]
      J091553+293326 & [300,300,300] & 0.68 & 42.9 & 0.05 & 23 & 15.71 & 18.45$_{-.16}^{+.19}$ & 0.08 & $8.50 \pm 0.16$ & $0.36_{-.02-.06}^{+.03+.07}$ \\ [.3ex]
      J095301$-$084034 & [300,300,300] & 0.78 & 33.9 & 0.92 & 34 & 15.29 & $>$ 18.71 & $<$ 0.04 & - & $>$ 0.40 \\ [.3ex]
      J1008+0621 & [300,300,300] & 0.73 & 46.0 & 0.05 & 26 & 17.68 & 19.80$_{-.13}^{+.14}$ & 0.14 & $14.8 \pm 4.32$ & $0.59_{-.03-.10}^{+.03+.11}$ \\ [.3ex]
      J103742+571158 & [300,300,300,300,300] & 0.76 & 51.8 & 0.01 & 20 & 15.97 & $>$ 19.95 & $<$ 0.03 & - & $>$ 0.62 \\ [.3ex]
      J105912$-$113424 & [300,300,300] & 0.85 & 42.7 & 0.05 & 21 & 15.86 & $>$ 19.08 & $<$ 0.05 & -  & $>$ 0.46 \\ [.3ex]
      J124313+362755 & [300,300,300,300] & 0.86 & 80.5 & 0.02 & 26 & 15.72 & 19.32$_{-.20}^{+.24}$ & 0.04 & - & $0.50_{-.04-.08}^{+.04+.10}$ \\ [.3ex]      
      J125311+530113 & [300,300,300,300,300] & 0.93 & 64.1 & 0.02 & 22 & 16.83 & 19.83$_{-.08}^{+.08}$ & 0.06 & $7.37 \pm 0.73$ & $0.59_{-.02-.10}^{+.02+.11}$ \\ [.3ex]
      J142700+234802 & [300,300,300] & 0.71 & 68.2 & 0.11 & 28 & 13.99 & 17.30$_{-.06}^{+.06}$ & 0.05 & - & $0.23_{-.01-.04}^{+.01+.05}$ \\ [.3ex]
      J144052+061033 & [300,300] & 0.98 & 56.6 & 0.07 & 26 & 16.89 & 19.59$_{-.10}^{+.11}$ & 0.08 & $7.24 \pm 2.31$ & $0.55_{-.02-.09}^{+.02+.10}$ \\ [.3ex]
      J154225+612950 & [300,300,300] & 1.02 & 45.4 & 0.03 & 33 & 14.98 & 18.68$_{-.17}^{+.20}$ & 0.03 & - & $0.39_{-.03-.07}^{+.03+.08}$ \\ [.3ex]
      J1624+5652 & [300,300,300] & 0.89 & 42.3 & 0.02 & 33 & 17.93 & 20.07$_{-.17}^{+.20}$ & 0.14 & - & $0.64_{-.04-.10}^{+.05+.12}$ \\ [.3ex]
      J1749+4321 & [300,300,300] & 1.09 & 29.2 & 0.07 & 30 & 17.41 & $>$ 20.73 & $<$ 0.05 & - & $>$ 0.79 \\ [.7ex]
      \hline
    \end{tabular}
  \end{center}
  \caption{\label{minimo_results} Summary of MiniMo results.  4+4 names are from CGRaBS \citep{het08}, 6+6 names 
are from CRATES \citep{het07}. No \{$\epsilon$, $\theta$\} was found with significance $>3\sigma$ for any BL Lac in these data.}
\end{sidewaystable}
\clearpage

\end{document}